\documentclass[showpacs,aps,prb,reprint,superscriptaddress]{revtex4-2}
\usepackage{bm}
\usepackage{graphicx}
\usepackage{graphics}
\usepackage{mathtools}
\usepackage{amsmath}
\usepackage{amsfonts}
\usepackage{amssymb}
\usepackage{epstopdf}
\usepackage{hyperref}
\usepackage{hyperref}
\hypersetup{
    colorlinks,%
    citecolor=blue,%
    linkcolor=blue,%
    urlcolor=blue
}
\usepackage{tikz}\usetikzlibrary{petri}
\usepackage{soul}
\usepackage[dvipsnames]{xcolor}

\begin{document}

\newcommand{\be}   {\begin{equation}}
\newcommand{\ee}   {\end{equation}}
\newcommand{\ba}   {\begin{eqnarray}}
\newcommand{\ea}   {\end{eqnarray}}
\newcommand{\ve}  {\varepsilon}

\title{ Onset of spin-valley order and Stoner boundaries in twisted WSe$_2$  }

\author{Lauro B. Braz}
\affiliation{
Instituto de F\'{\i}sica, Universidade de S\~ao Paulo, Rua do Mat\~ao 1371, S\~ao Paulo, S\~ao Paulo 05508-090, Brazil
}

\author{Luis G.~G.~V. Dias da Silva}
\affiliation{
Instituto de F\'{\i}sica, Universidade de S\~ao Paulo, Rua do Mat\~ao 1371, S\~ao Paulo, S\~ao Paulo 05508-090, Brazil
}

\date{ \today }

\begin{abstract}
We investigate spin–valley instabilities and their connection to the 
magnetically ordered states
recently observed in the twisted bilayer dichalcogenide WSe$_2$ at a $5^o$ twist angle. 
 Starting from an effective three-orbital faithful Wannier model for the spin-locked moiré bands, combined with orbital-dependent Hubbard interactions, we analyze the evolution of magnetic instabilities as a function of carrier density using the matrix random phase approximation (mRPA) approach. By computing  the Stoner boundary lines from the spin-valley susceptibilities over  the electric-field by hole filling phase diagram, we show that the spin–valley instabilities result in ordered states in the region close to the Lifshitz transition at the topmost moiré valence band, marked by crossing of the Van Hove singularity in the density of states. These spin-valley ordered states are dominated by interorbital spin–valley–flips involving the $MM$ and $MX$ moiré orbitals and occur at different momenta in each side of the Van Hove line, indicating a distinct spatial dependence of the spin-valley order parameter depending on the hole filling. Moreover, the corresponding Stoner boundaries exhibit strong fluctuations on its flanks, which can favor superconducting states in the regions close to the spin-valley-ordered ones. This mechanism provides a natural description for a reentrant superconducting dome consistent with the experimental results. As such, our results suggest spin–valley fluctuations near the van-Hove line as the microscopic origin of the reentrant superconductivity in twisted WSe$_2$.
\end{abstract} 

\maketitle

\section{Introduction}
\label{sec:Intro}

The discovery of correlated and superconducting states in moiré materials  \cite{AndreiMarvelsOfMoiréMaterials2021} has opened a different avenue for exploring strongly interacting two-dimensional (2D) systems in a regime where electronic bandwidths, interaction strengths, and topology can all be tuned by design.
Since the first observations of correlated insulators and superconductivity in twisted bilayer graphene \cite{cao_unconventional_2018}, moiré engineering has emerged as a powerful platform for realizing flat electronic bands with enhanced Coulomb correlations and tunable many-body ground states.

Beyond graphene, transition-metal dichalcogenide (TMD) bilayer systems 
\cite{Li_NatScienceRev_2026}
have recently attracted considerable attention, as they naturally host large spin–orbit coupling, valley-contrasting physics, and strong dielectric confinement, providing a qualitatively distinct route towards correlated and topological quantum phases \cite{wangCorrelatedElectronic2020,xuTunableBilayerHubbard2022,andersonCorrelatedMagnetic2023,caiFractionalQuantumHall2023,zengFractionalChernInsulator2023,parkFractionalyQuantized2023,FouttyMappingTwistTunedMultiband2024,XuObservationIntegerFractionalQuantumAnomalous2023,GhiottoStonerInstabilitiesIsing2024,GhiottoQuantumCriticalityTwisted2021, KielyContinuousWigner-MottTransitions2024}.

Among TMD moiré materials, twisted bilayer WSe$_2$ (tWSe$_2$) has recently emerged as a paradigmatic system for correlated electron physics in flat bands. Recent experiments \cite{Xia_Nature_833838_2025,Guo_Nature_839845_2025,Knuppel_NatComm_2025,Xia_BandwidthTunedMottTransitionSuperconductivityMoireWSe2_2026,Guo_AngleEvolutionSuperconductingPhaseDiagram_tWSe2_Nature_2026} have provided compelling evidence for superconductivity in tWSe$_2$ at twist angles in the $3^o - 5^o$ range. These superconducting domes typically appear in narrow ranges of carrier density and displacement field, reminiscent of the behavior seen in magic-angle graphene, yet with distinct symmetry and spin properties stemming from the spin–valley–locked nature of TMDs. 

Although superconducting transitions with critical temperatures of a few hundred millikelvin have been observed near integer fillings of the moiré bands in both cases, at lower twist angles ($\sim3^o$), superconductivity is found in close proximity to a correlated insulating state \cite{Xia_Nature_833838_2025,Xia_BandwidthTunedMottTransitionSuperconductivityMoireWSe2_2026,Guo_AngleEvolutionSuperconductingPhaseDiagram_tWSe2_Nature_2026}, resembling the strong-coupling physics of twisted bilayer graphene. By contrast, at higher twist angles ($\sim5^o$), superconductivity was found close to a Fermi-surface-reconstructed magnetic state, which remains metallic \cite{Guo_Nature_839845_2025,Xia_BandwidthTunedMottTransitionSuperconductivityMoireWSe2_2026,Guo_AngleEvolutionSuperconductingPhaseDiagram_tWSe2_Nature_2026}. More intriguingly, the experiment at $5^o$ angle \cite{Guo_Nature_839845_2025,Xia_BandwidthTunedMottTransitionSuperconductivityMoireWSe2_2026,Guo_AngleEvolutionSuperconductingPhaseDiagram_tWSe2_Nature_2026} found two distinct superconducting regimes: First, along the Van Hove singularity and near $\nu=-1$ electron filling, superconductivity emerges as an isolated superconducting dome. At high electric field, when the density of states becomes even more pronounced, a magnetic state is triggered and splits the superconducting dome in two, a characteristic of a reentrant superconducting state.

From a theoretical point of view, the superconducting mechanism in tWSe$_2$ remains subject to intense debate. Theoretical and experimental studies have established that tWSe$_2$ realizes narrow, isolated moiré bands with strong spin–valley locking and significant Berry curvature.
Since the underlying spin–orbit coupling breaks spin degeneracy, the resulting low-energy electronic structure can be viewed as a set of spin-valley polarized bands with effective triangular-lattice geometry, where local Coulomb interactions compete with itinerancy and topological band properties \cite{PanBandTopologyHubbard2020,Kormányos_kpTheoryTwoDimensional2015,DevakulMagicInTwistedTransition2021,CrepelBridgingSmallLargeTwisted2024,ZhangUniversalMoire-Model-Building2024}.

The moiré bands of tWSe$_2$ can be well described by an effective Hubbard model on a triangular lattice, capturing the interplay between strong on-site and longer-range Coulomb interactions and the geometric frustration inherent to the lattice \cite{PanBandTopologyHubbard2020}.
Due to the large spin–orbit coupling, the spin and valley degrees of freedom are intertwined, reducing the effective symmetry and allowing for unconventional pairing channels, including mixed singlet–triplet or chiral states \cite{AkbarTopologicalSuperconductivityMixed2024,ZegrodnikMixedSinglet-TripletSuperconductingState2023,WuPair-Density-WaveChiralSuperconductivity2023}.

Early studies based on mean-field and strong-coupling approaches have found evidence for correlated insulators and possible magnetically mediated superconductivity \cite{KimTheoryCorrelatedInsulatorsSuperconductor2025,AbouelkomsanMultiferroicityTopologyTwisted2024}.
In addition, theoretical works have emphasized the relevance of Van Hove singularities tunable by twist angle or displacement field, which can enhance interaction-driven instabilities in specific momentum channels \cite{HsuSpin-valleyLockedInstabilities2021,FouttyMappingTwistTunedMultiband2024,ZhuSuperconductivityInTwistedTransition2025}.

A variety of complementary theoretical frameworks have been developed to analyze the interplay of correlations and superconductivity in tWSe$_2$.
Continuum-model and Wannier-based approaches have clarified the nature of the moiré bands and their evolution with the twist angle, linking small-angle and large-angle regimes through realistic three-band tight-binding models \cite{CrepelBridgingSmallLargeTwisted2024,CrépelMillisSpinonPairingInduced2024}.
Strong-coupling and cluster DMFT studies have demonstrated the emergence of $d+id$-wave superconductivity in related TMD-based triangular lattices \cite{BelangerSuperconductivityTwistedBilayerTransitionMetalDichalcogenideQuantumCluster2022}, while renormalization-group and variational studies have identified competing spin-singlet, triplet, and pair-density-wave (PDW) instabilities \cite{ChenSingletTripletPairDensityWave2023,WuPair-Density-WaveChiralSuperconductivity2023,ZhouModelValleyContrastingFlux2023,Fischer_Phys.Rev.X_041055_2025}.
More recent theoretical efforts have focused on the possibility of topological superconductivity emerging from repulsive interactions or from valley-contrasting fluxes \cite{TuoTheoryTopologicalSuperconductivityAntiferromagnetic2025,GuerciTopologicalSuperconductivityRepulsive2024,ZegrodnikMixedSinglet-TripletSuperconductingState2023,AkbarTopologicalSuperconductivityMixed2024,XieOrbitalFuldeFerrellPairing2023}.
Despite this growing theoretical landscape, the microscopic mechanism responsible for pairing and the symmetry of the superconducting order parameter remains unsettled. In addition, several mean-field \cite{CrépelMillisSpinonPairingInduced2024,KimTheoryCorrelatedInsulatorsSuperconductor2025,MunozSegovia_Phys.Rev.B_085111_2025}, renormalization group \cite{HsuSpin-valleyLockedInstabilities2021,Fischer_Phys.Rev.X_041055_2025}, and RPA \cite{SchradeFuNematicChiralTopologicalSuperconductivity2024} studies have proposed spin-valley order in tTMDs. 

Despite the recent activity in the field, a detailed  {theoretical understanding of the magnetic order} observed in the $\sim5^o$ experiments \cite{Guo_Nature_839845_2025} is still absent. 
In this work, we aim to fill this gap by providing a theoretical study of this topic based on the matrix-RPA (mRPA) method \cite{graserNeardegeneracySeveralPairing2009,altmeyerRoleVertexCorrections2016}. 
Within this context, the mRPA framework provides a natural framework to investigate the leading instabilities in tWSe$_2$ from a weak-to-intermediate coupling perspective, accounting for pairing vertex diagrams beyond what is usually known as RPA~\cite{esirgenMathitWavePairing1999a,altmeyerRoleVertexCorrections2016}. 
In fact, one of the results reported here is that this approach naturally accounts for a signature of reentrant superconductivity resulting from a Stoner-like instability \cite{Braz:Phys.Rev.B:184502:2024} along the van-Hove singularity line.

This paper is organized as follows: the three-orbital Wannier-based model for tWSe$_2$ is discussed in Sec.~\ref{sec:modelmethods} while the details of the matrix-RPA calculations are given in Sec.~\ref{sec:RPAformalism}. The main results for the spin-valley instabilities and the Stoner boundary are given in \ref{sec:competition}. Our overall conclusions are summarized in Sec.~\ref{sec:Conclusions}.

\section{Wannier-based model}
\label{sec:modelmethods}

\begin{figure}[t]
\begin{center}
\includegraphics[width=1.0\columnwidth]{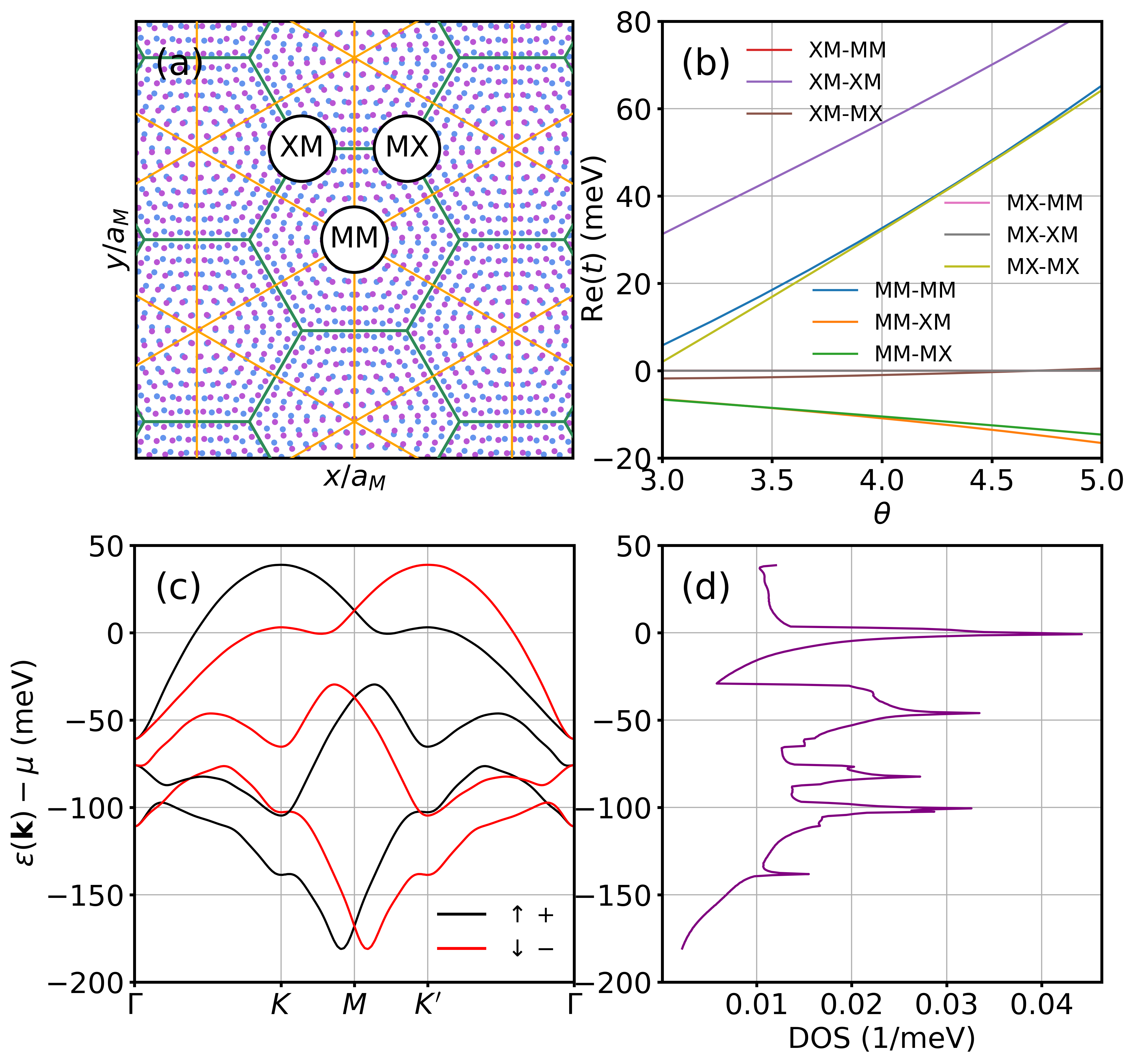}
\caption{Schematic of the non-interacting model we used, 
	obtained by Wannierizing the tTMDh continuum model~\cite{CrepelBridgingSmallLargeTwisted2024}.
	(a) shows a depiction of the real-space moir\'e superlattice for $\theta=5^o$  centered at an $MM$ sublattice site.
    This model considers two lattices, a triangular lattice of $MM$ sites (yellow lines), and a hexagonal lattice of $MX$ and $XM$ sites (green lines). Intralattice and interlattice hoppings are taken into account for distances up to $9$ unit cells.
    (b) shows the nearest-neighbor hoppings and onsite energies used in our model for electric field $E_z=20$ meV as a function of twist angle.
    In (c), we show a representative band structure of the three-orbital model for electron filling $\nu=-1$ and electric field $E_z=20$ meV, whereas (d) shows the respective energy-dependent total density of states.
}
\label{fig:model}
\end{center}
\end{figure}

\subsection{Effective Hamiltonian}
\label{sec:effectiveHami}

An effective interacting model for  {hole-doped} tWSe$_2$ at a 5$^o$ twist angle can be represented by the following  Hamiltonian
\begin{equation}
    H = H_0 + H_\text{int},
\end{equation} 
where $H_0$ is the non-interacting term given by the three-orbital model  {constructed from spin-valley locked bands \cite{Xiao_Phys.Rev.Lett._196802_2012} from both monolayers presented in }  Refs.~\cite{CrepelBridgingSmallLargeTwisted2024,Fischer_Phys.Rev.X_041055_2025} 
\begin{equation}
    H_0 = \sum_{\xi\boldsymbol{R}\boldsymbol{R}'}\sum_{pp'}t^\xi_{\boldsymbol{R}p,\boldsymbol{R}'p'}c_{\boldsymbol{R} p \xi}^{\dagger}c_{\boldsymbol{R}' p' \xi},
    \label{eq:TBmodel}
\end{equation} 
where $c_{\boldsymbol{R} p \xi \sigma}^{\dagger}$ creates an 
electron with spin-valley  {$\xi=\{(\uparrow+),(\downarrow-\})$}, in the Wannier state $|\boldsymbol{R}, p \rangle$ 
centered at position $p=\{MM,XM,MX\}$, as represented in Fig.~\ref{fig:model}(a), in the unit cell located at the 
moir\'e lattice vector $\boldsymbol{R}$.  {The hopping matrix includes} intersublattice and intrasublattice hoppings up to nine unit-cell distances.
 {
We show in Fig.~\ref{fig:model}(b) the onsite energies and nearest-neighbor hoppings for each orbital in the model for fixed $E_z=20$ meV and as a function of twist angle $\theta$.
The electron occupation $n$ is enforced by integrating the density of states from the bottom of the lower band to a fixed-energy chemical potential $\mu$.
}

The interacting part of the Hamiltonian, $H_\text{int}$, is given by \cite{CrepelBridgingSmallLargeTwisted2024}
\begin{equation}
H_{\text{int}} = \sum_{\boldsymbol{R}p\xi} U_{p} \hat{N}_{\boldsymbol{R} p \xi} \hat{N}_{\boldsymbol{R} p {(-\xi)}},
\label{Hint}
\end{equation} 
where $\hat{N}_{\boldsymbol{R} p \xi}=c_{\boldsymbol{R} p \xi }^{\dagger} c_{\boldsymbol{R} p \xi} $ is 
the number operator related to the Wannier state.
    Here, we consider $U_{XM}=U_{MX}\approx41.3$ meV and $U_{MM}/U_{MX}\approx0.90$,  {which reproduces the onset  near $\nu=-1$ seen in experiments. , which here occurs  for electric fields $E_z\sim20$ meV.
 The robustness of our results to changes on the interaction parameters is demonstrated in Appendix~\ref{appendix:robust}.

\subsection{Density of states: Van Hove singularity and Lifshitz transition}
\label{sec:bandstructure}

Let us first review the main features of the non-interacting band structure given by diagonalizing $H_0$ (Eq. \eqref{eq:TBmodel}). As shown in Fig.~\ref{fig:model}(a), the three orbitals represent a triangular ($MM$ orbital) and hexagonal ($XM$ and $MX$ orbitals) lattices. Also, since the strong spin-orbit coupling in  {each WSe$_2$} monolayer induces spin-valley locking \cite{Kormányos_kpTheoryTwoDimensional2015}, spin and valley  {quantum numbers} are denoted by  {a single spin-valley index} $\xi=\{(\uparrow+),(\downarrow-)\}$. Time-reversal symmetry (momentum inversion plus $\xi\rightarrow-\xi$)  ensures that $H_0$ is block-diagonal in the spin-valley index $\xi$. 

The three-orbital construction of Ref. \cite{Fischer_Phys.Rev.X_041055_2025} takes into account the topological properties of the low-lying hole bands in the continuum model, providing an overall zero net Chern number, thus allowing for a Wannier obstruction-free, faithful description of the band structure.  We show a representative band structure in Fig.~\ref{fig:model}(c) for electric field $E_z=20$ meV and electron filling $\nu=-1$ and the respective density of states in Fig.~\ref{fig:model}(d). 
Here, the filling factor is defined by normalizing the three-band total occupation $n$ to $6$ electrons in total, three for each spin-valley such that $\nu=n-6$.
The topmost band will be the hole-doping band, which is experimentally accessible and hosts up to two holes ($\nu=-2$).

In Fig.~\ref{fig:dos}(a) we show a color map of the electron density of states 
as a function of the displacement field $E_z$ and the electron filling factor $\nu$, which can be directly compared with the resistance maps in the $\sim5^o$ twist angle experiment \cite{Guo_Nature_839845_2025,Xia_BandwidthTunedMottTransitionSuperconductivityMoireWSe2_2026,Guo_AngleEvolutionSuperconductingPhaseDiagram_tWSe2_Nature_2026}. A darker green  color marks the Van Hove singularity.

In order to properly compare our results to the experimental phase diagram, we notice that Refs.~\cite{Guo_Nature_839845_2025,Xia_BandwidthTunedMottTransitionSuperconductivityMoireWSe2_2026, Guo_AngleEvolutionSuperconductingPhaseDiagram_tWSe2_Nature_2026} report superconductivity near the Van Hove singularity at $\nu=-1$ for $5^o$ twist angle.
This restricts the relevant electric field regions to about $E_z\sim20$ meV.
Figure~\ref{fig:dos}(b) shows  {(from top to bottom)} the Fermi surfaces  {at constant $E_z=20$ meV and filling factors $\nu=-1.1, -1.05, -095$, respectively}  marked by circle (yellow), square (green) and  {diamond} 
(purple) symbols in Fig.~\ref{fig:dos}(a).
 {This sequence nicely illustrates the Lifshitz transition marked by the appearance of a small hole pocket at $K^\prime$ and the subsequent crossing of the Van Hove singularity (VHS), leading to a sharp increase in the density of states \cite{MunozSegovia_Phys.Rev.B_085111_2025}.}
 {Consistently, Fig.~\ref{fig:model}(b) shows that at $E_z=20$ meV (high electric field), the $MX-XM$ hoppings are nearly zero for $5^o$ twist angle, which aligns with the high layer polarization picture at this electric field range.
In conclusion, the electric field $E_z$ pushes the $XM$ orbital is away from the Fermi level.}

\begin{figure}[t]
\begin{center}
\includegraphics[width=1.0\columnwidth]{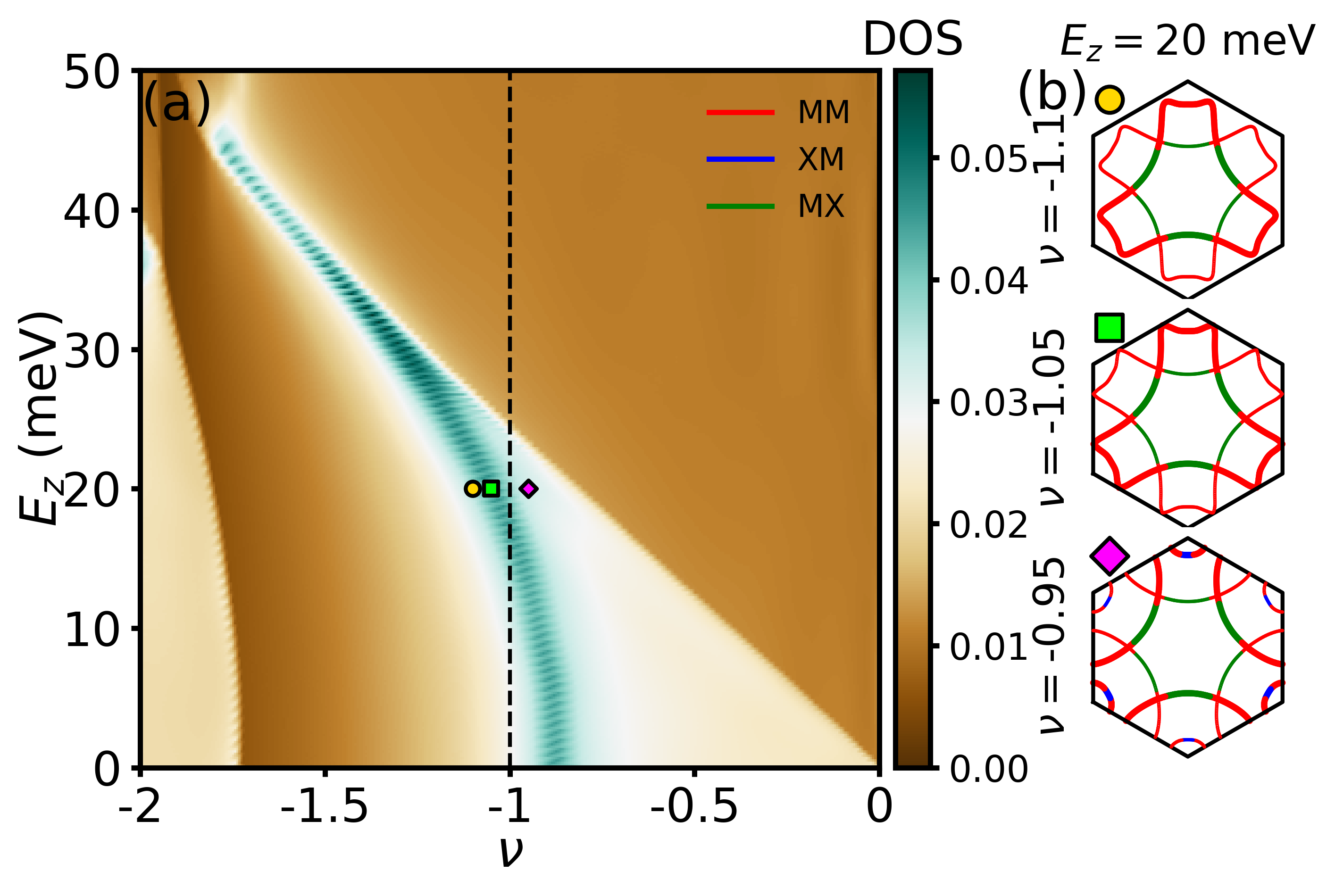}
	\caption{ (a) shows a color map of the density of states in an electric field $E_z$ by electron filling $\nu$ diagram.
    A black vertical dashed line marks the $\nu=-1$ filling, which is relevant for the experimental phase diagram.
    The  {dark green}  region of the phase diagram marks the VHS.
    The yellow circle, green square and purple  {diamond} symbols marks points of the phase diagram at constant electric field $E_z=20$ meV for which we show Fermi surfaces in (b).
    From top to bottom, (b) shows a Lifshitz transition and the crossing of the VHS by the Fermi energy, marked by a steep increase in the DOS. 
    Orbital contributions to the Fermi surfaces are shown in colors, i.e. red for $MM$, blue for $XM$ and green for $MX$.
    Spin-valleys $(\uparrow+)$ and $(\downarrow-)$ are represented by thicker and thinner lines, respectively.
}
\label{fig:dos}
\end{center}
\end{figure}

\section{Spin-valley fluctuations: matrix-RPA formalism}
\label{sec:RPAformalism}
 We study the Stoner phase diagram of $5^o$-twist WSe$_2$ in the weak-to-intermediate-coupling limit using the matrix RPA.
We remark that at electric field $E_z=20$ meV, the bandwidth of the three-orbital model is $W\approx200$ meV, whereas the largest interaction strength used throughout this work is $U_{XM}=U_{MX} \approx 40$ meV, resulting in a ratio $U_{MX}/W\approx0.2$, which naturally places the system in the weak-to-intermediate-coupling limit.
In fact, the tTMDs show a tendency to more itinerant behavior at higher twist angles \cite{CrépelMillisSpinonPairingInduced2024}.
Moreover, a systematic comparison of the weak-coupling functional renormalization group approach to several twist-angle experiments suggests that the weak-coupling limit is enough to roughly describe the experimental phase diagrams \cite{Guo_AngleEvolutionSuperconductingPhaseDiagram_tWSe2_Nature_2026}.

In the following, we describe the matrix RPA steps used in our analysis. 

Our goal is to probe the instabilities caused by spin-valley quantum fluctuations. By incorporating the spin and charge fluctuations arising from Coulomb interactions on the moiré lattice, the mRPA approach allows one to identify the dominant momentum-dependent pairing channels and to distinguish between competing spin-singlet and spin-triplet superconducting states.  

The spin–valley–locked band structure of tWSe$_2$ introduces additional complexity, as spin and momentum are no longer independent quantum numbers.
Consequently, the effective interaction vertices acquire a pronounced dependence on valley and orbital composition, which can qualitatively modify the competition between even- and odd-parity pairing.
This formalism has been successfully applied in related 2D systems to describe nematic, chiral, and topological superconductivity mediated by spin fluctuations \cite{SchradeFuNematicChiralTopologicalSuperconductivity2024,MunozSegovia_Phys.Rev.B_085111_2025}.
In the context of moiré TMDs, however, a fully microscopic RPA treatment that explicitly includes spin–valley locking, realistic band dispersions, and interlayer coupling has not yet been comprehensively explored.

The starting point is the bare (non-interacting)  {static} multi-orbital susceptibility matrix elements \cite{graserNeardegeneracySeveralPairing2009,ParcolletTRIQSToolboxForResearch2015,StrandTPRFToolboxTRIQS2019,Braz:Phys.Rev.B:184502:2024}  {relating orbital ($p,q,r,t$) and spin-valley ($\xi, \xi^{\prime})$ indices as:} 
\begin{equation}
\begin{split}
    \left[\chi_{0}(\boldsymbol{q})\right]_{r\xi , t \xi^\prime}^{p \xi^\prime, q \xi}=-\frac{T}{N_{\boldsymbol{k}}}&\sum_{\boldsymbol{k}i\omega_n}G_{t\xi p\xi}(\boldsymbol{k},i\omega_n)\\
    &\times G_{q\xi'r\xi^\prime}(\boldsymbol{k}+\boldsymbol{q},i\omega_n),
\end{split}
\label{eq:chi_0td}
\end{equation}
 {where
\begin{equation}
    G_{t\xi p\xi}(\boldsymbol{k},i\omega_n)=\sum_{\nu} \frac{\psi_{\nu}^{t \xi}(\boldsymbol{k})\psi_{\nu}^{p \xi*}(\boldsymbol{k})}{i\omega_n-E_{\nu}^{\xi}(\boldsymbol{k})},
\label{eq:GF}
\end{equation}
}
which depend on the eigenvalues $E^{\xi}_\nu (\boldsymbol{k})$ of the non-interacting Hamiltonian 
$H_0$ [Eq.~\eqref{eq:TBmodel}], and on the eigenvector coefficients 
{$\psi_{\nu}^{p \xi}(\boldsymbol{k}) \equiv \langle p | \nu \boldsymbol{k} \rangle_\xi$}, 
which correspond to the projection of band state {$| \nu \boldsymbol{k} \rangle_{\xi}$} into the Wannier orbital {$|p \rangle_\xi$}  at valley $\xi$.

In our model with three Wannier orbitals per spin-valley, $\hat{\chi}_{0}(\boldsymbol{q})$ is a {$36 \times 36$} matrix 
{spanning  the $\left\{p \xi', q \xi \right\}$ basis. In Eq. \eqref{eq:chi_0td}, }
$N_{\boldsymbol{k}}$ is the number of Brillouin zone $\boldsymbol{k}$-points considered in the summation, and 
$T$ is the temperature. Throughout this work, we used a summation grid of 
$256\times256$ $\boldsymbol{k}$-points in the triangular lattice Brillouin zone, up to $\pm 300$ meV imaginary frequencies using the discrete Lehmann representation \cite{ParcolletTRIQSToolboxForResearch2015,StrandTPRFToolboxTRIQS2019,KayeDiscreteLehmannRepresentationImaginary2022} at a temperature $T=0.03\;\text{meV}=352\;\text{mK}$.
 { Further numerical details on the implementation of the bare susceptibilities are discussed in Appendix~\ref{appendix:efficiency}. }

We write the RPA spin-valley susceptibility suitable for detecting magnetism 
and valley order in the system as \cite{graserNeardegeneracySeveralPairing2009,RoemerKnightShiftLeadingSuperconductingInstability2019}
 {
\begin{align}           
\hat{\chi}(\boldsymbol{q})	=\hat{\chi}_{0}(\boldsymbol{q})\left[\hat{1}-\hat{U}\hat{\chi}_{0}(\boldsymbol{q})\right]^{-1}, \label{eq:chi}
\end{align} 
}
where the non-zero $\hat{U}$ matrix elements can be written in terms of the $U_p$ defined in Eq.~\eqref{Hint} as: 
\begin{align}
\left[ U\right]^{p \xi , p\xi' }_{p \xi , p\xi' } &= U_{p} \delta_{\xi \xi'}.
\label{eq:U}
\end{align}

The generalized Stoner criterion [namely, the vanishing of the denominator in Eq.~\eqref{eq:chi}] establishes the condition for the transition between a paramagnetic (uniform density) state possibly favoring the SC phase, and a spin-valley ordered one, which necessarily suppresses superconductivity. 
We define the spin-valley ($\alpha$) critical Stoner parameter by solving the 
following eigenvalue equations~\cite{sakakibaraOriginMaterialDependence2012} 
\begin{equation}
\begin{split}
    & \mbox{det}\left( \hat{1}\alpha - \hat{U}\hat{\chi}_0 \right) = 0.
\end{split}
\label{eq:stoner}
\end{equation}
Defining the main Stoner parameter $\max \{ \alpha \} \equiv \alpha_c$, the Stoner criterion is fulfilled when $\alpha_c =1$.  Notice that $\alpha_c > 1$ indicates spin-valley ordering for any nonzero value of the interaction $U_p$, thus completely suppressing superconductivity. As such, we define the line in parameter space marked by $\alpha_c =1$ as the \emph{Stoner boundary} \cite{Braz:Phys.Rev.B:184502:2024}.

 In this sense, the diverging components of the RPA spin-valley susceptibility matrix $\hat{\chi}(\boldsymbol{q})$ [Eq.~\eqref{eq:chi}] at $\boldsymbol{q}=\boldsymbol{Q}$ and at a critical parameter can be associated with a spin-valley phase transition and the emergence of an order parameter $\boldsymbol{\Delta}(\boldsymbol{Q})$ with ordering vector $\boldsymbol{Q}$. 

 In fact, as discussed in Refs. \cite{boehnkeSusceptibilitiesMaterialsMultiple2015,BrazCompetingMagneticStatesSurfaceMultilayer2024}, information on the order parameter at the instability point ($\alpha_c=1$) can be obtained from the eigenvector $\Delta_{p\xi t\xi^\prime}$ associated with the leading eigenvalue $\chi^{\rm eig}(\boldsymbol{Q})$ of $\hat{\chi}(\boldsymbol{Q})$: 
\begin{equation}
   \sum_{r\xi , q \xi^\prime} \left[\hat{\chi}(\boldsymbol{Q})\right]_{ r\xi , q \xi^\prime}^{p\xi t\xi^\prime  } \Delta_{ r\xi  q \xi^\prime} = \chi^{\rm eig}(\boldsymbol{Q})  \Delta_{p\xi t\xi^\prime}
\label{eq:EigenvalueChi}
\end{equation}
 {By finding the vector $\boldsymbol{Q}$ at which $\chi^{\rm eig}(\boldsymbol{q}=\boldsymbol{Q})$ diverges for $\alpha_c=1$, the order parameter can then be written as \cite{boehnkeSusceptibilitiesMaterialsMultiple2015} }
\begin{equation}
    \boldsymbol{\Delta}(\boldsymbol{Q})\propto \sum_{p\xi s\xi^\prime}\Delta_{p\xi t\xi^\prime}\langle\boldsymbol{S}_{p\xi s\xi^\prime}(\boldsymbol{Q})\rangle ,
\label{eq:OP}
\end{equation}
where
\begin{equation}
    \boldsymbol{S}_{p\xi s\xi'}(\boldsymbol{Q})=\sum_{\boldsymbol{k}} c_{(\boldsymbol{k}+\boldsymbol{Q})p\xi}^{\dagger}\left(\boldsymbol{\sigma}\right)_{\xi\xi'}c_{\boldsymbol{k}s\xi'}.
\label{eq:spin}
\end{equation}
is the spin operator and 
$\boldsymbol{\sigma}=(\sigma_x, \sigma_y, \sigma_z)$ is the vector of Pauli matrices in the spin-valley basis. Notice that the $z$ direction of $\boldsymbol{\Delta}(\boldsymbol{Q})$ corresponds to the direction of the spin-valley polarization.
 Thus, we can obtain information on the order parameter $\boldsymbol{\Delta}(\boldsymbol{Q})$  by numerically evaluating Eq.~\eqref{eq:EigenvalueChi} near/at the Stoner boundary. 
%

\begin{figure}[t]
\begin{center}
\includegraphics[width=1.0\columnwidth]{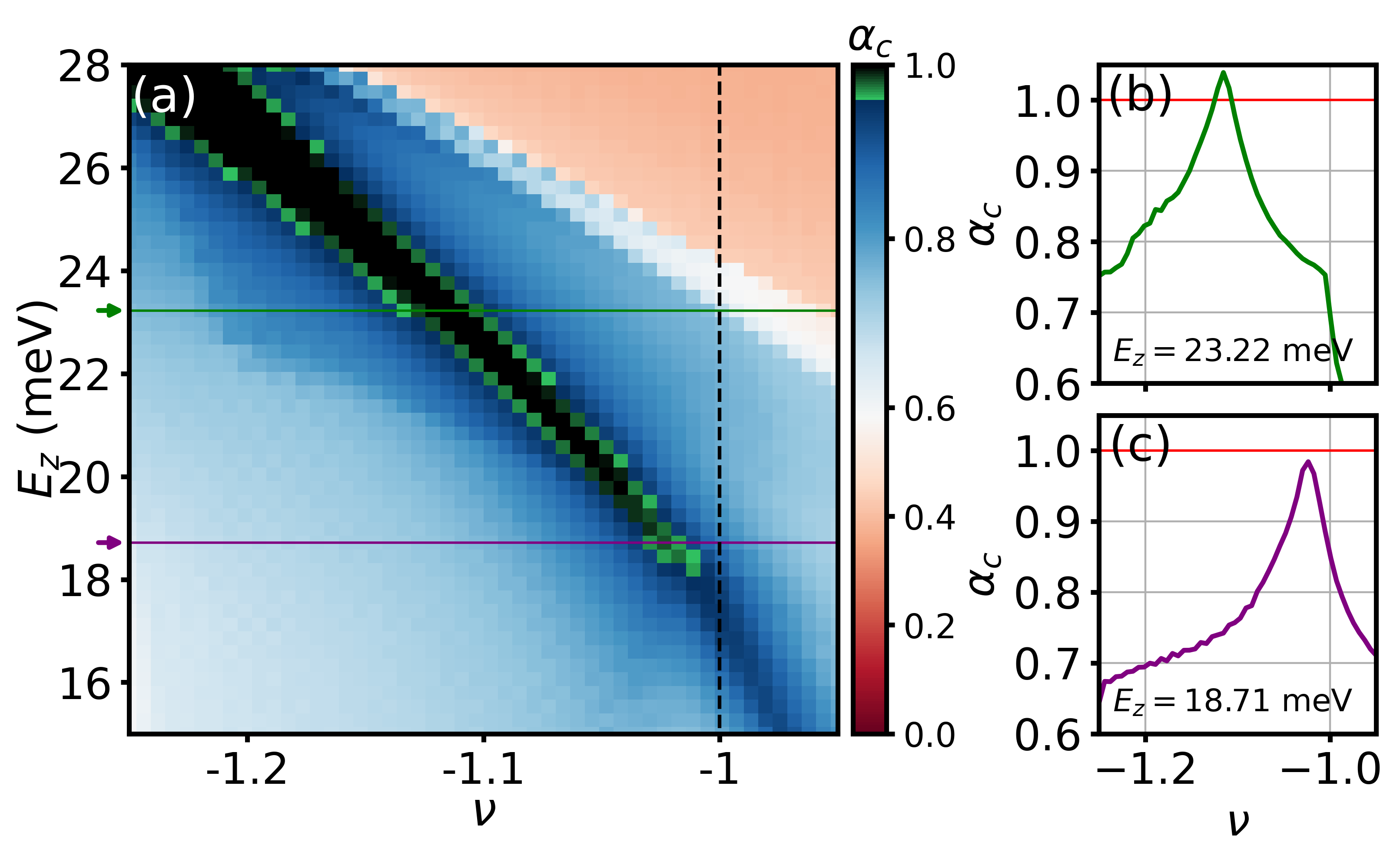}
\caption{
	Stoner phase diagram.  (a) shows the critical Stoner parameter $\alpha_c$ in the electric field $E_z$ by electron filling $\nu$ diagram at Hubbard interaction $U_{XM}=U_{MX}\approx41.3$ meV and $U_{MM}/U_{XM}\approx0.90$.
    The green color interpolates with black for $0.95\leq\alpha_c\leq1$. The black colors characterizes $\alpha_c\geq1$,  {which
    indicates that the Stoner criterion, marking the transition to an spin-valley ordered state, has been met}.
     (b) and (c) show constant electric field cuts for $E_z=23.22$ meV (green) and $E_z=18.71$ meV (purple) of the phase diagram.
    The constant $\alpha_c=1$ red line marks the Stoner boundary.
}
\label{fig:stoner}
\end{center}
\end{figure}

\section{Spin-valley ordered phase}
\label{sec:competition}
We now turn to the phase diagram  {based on the Stoner criteria for  spin-valley fluctuations, showing the boundary  marking the spin-valley ordered phase. 

\subsection{Stoner boundary}
\label{sec:stoner_boundary}

Figure~\ref{fig:stoner}(a) shows the critical Stoner parameter $\alpha_c$  {in a colormap as a function of the displacement field $E_z$ and }   electron filling $\nu$.
The green to black gradient marks the $0.95\leq\alpha_c\leq1$ region.
Black color denotes $\alpha_c\geq1$, defining the spin-valley ordered region.

At $E_z\approx20$ meV, there is spin-valley order along the Van Hove singularity line (black color), and high $\alpha_c$ regions at the flanks of the ordered state.
Fig.~\ref{fig:stoner}(b) shows an $E_z=23.07$ meV cut [green arrow in Fig.~\ref{fig:stoner}(a)], where we remark the ordered region ($\alpha_c>1$).
When $E_z\approx18$ meV, there is no spin-valley order, but the critical Stoner parameter is still very high, as depicted in Fig.~\ref{fig:stoner}(c) [purple arrow in Fig.~\ref{fig:stoner}(a)].

 {The results shown in Fig.~\ref{fig:stoner} are consistent with the reentrant superconducting state as observed in the experiments \cite{Guo_Nature_839845_2025} and constitute one of the main results of this paper. Since the superconducting phase is favored near the instability line, where the pairing is stronger and the SC order parameter tends to be larger \cite{graserNeardegeneracySeveralPairing2009, Braz:Phys.Rev.B:184502:2024}, } the green-black-green color pattern  { shows that the system can go from a superconducting-ordered state (green) to a spin-valley ordered state (black)} by varying the electron filling in the region $20\lesssim E_z\lesssim21$ meV shown in Fig.~\ref{fig:stoner}(a),  {thus characterizing a reentrant superconducting state similar to that seen in the experimental results of Refs.~\cite{Guo_Nature_839845_2025,Xia_BandwidthTunedMottTransitionSuperconductivityMoireWSe2_2026,Guo_AngleEvolutionSuperconductingPhaseDiagram_tWSe2_Nature_2026}.} 


\subsection{Spin-valley order and the connection to reentrant superconductivity}
\label{sec:spin_valley}

In order to obtain the spin-valley order parameter $\boldsymbol{\Delta}(\boldsymbol{Q})$ [Eq.~\eqref{eq:OP}], we numerically solve Eq.~\eqref{eq:EigenvalueChi} near the $\alpha_c\!=\!1$ transition lines in the phase diagram shown in  Fig.~\ref{fig:stoner}, to determine the leading eigenvalues $\chi^{\rm eig}(\boldsymbol{q})$  of the susceptibility matrix, as well as the divergent nesting vectors $\boldsymbol{Q}$. These results are shown in  Fig.~\ref{fig:chi}, where we show $\chi^{\rm eig}(\boldsymbol{q})$ close  to the Stoner boundary (we set $U=0.99U_c$) for $E_z=20$ meV and fillings $\nu$ corresponding to  the colored markers shown in Fig.~\ref{fig:dos}(a).

For $\nu=-1.05$, corresponding to the Van Hove line in the DOS, $\chi^{\rm eig}(\boldsymbol{q})$ diverges at $\boldsymbol{Q}\equiv\boldsymbol{Q}_1=0$. In fact, the $\boldsymbol{Q}=0$ divergence persists for the \emph{entire} Van Hove line in the phase diagram. This result can be understood as follows:  
at $\boldsymbol{q}=0$ and $T=0$, the homogeneous non-interacting susceptibility 
is proportional to the density of states at the Fermi level \cite{graserNeardegeneracySeveralPairing2009}.
 {As such, } a divergence in the density of states  (such as a Van Hove singularity associated with a Lifshitz transition) will manifest itself as a  $\boldsymbol{Q}=0$ divergence in both the homogeneous and interacting susceptibilities.

Given the results for $\chi^{\rm eig}(\boldsymbol{q})$ and the corresponding eigenvectors at the Stoner boundary, the leading contributions to the spin-valley order parameter can be obtained from Eq.~\eqref{eq:OP}.  
 { We detail the procedure to extract the main order parameter components in Appendix~\ref{appendix:op}. }
For $\nu=-1.05$ and $\nu=-0.95$, we find that the leading eigenvector components are $\Delta_{(MM\uparrow)(MM\uparrow)}=-\Delta_{(MX\downarrow)(MX\downarrow)}$. Thus, our results are consistent with a spin-valley order parameter of the form 
\begin{equation}
\begin{split}
    \Delta_z(\boldsymbol{Q} \approx \boldsymbol{0})\propto &\Delta\Big[ \sum_{\boldsymbol{k}}\langle c_{(\boldsymbol{k})(MM\uparrow)}^{\dagger}c_{\boldsymbol{k}(MM\uparrow)}\rangle \\   
    &+\langle c_{(\boldsymbol{k})(MX\downarrow)}^{\dagger}c_{\boldsymbol{k}(MX\downarrow)}\rangle\Big],
\end{split}
\label{eq:OP_result}
\end{equation}
where we have defined $\Delta \equiv \Delta_{(MM\uparrow)(MM\uparrow)}$. 
\begin{figure}[t]
\begin{center}
\includegraphics[width=1.0\columnwidth]{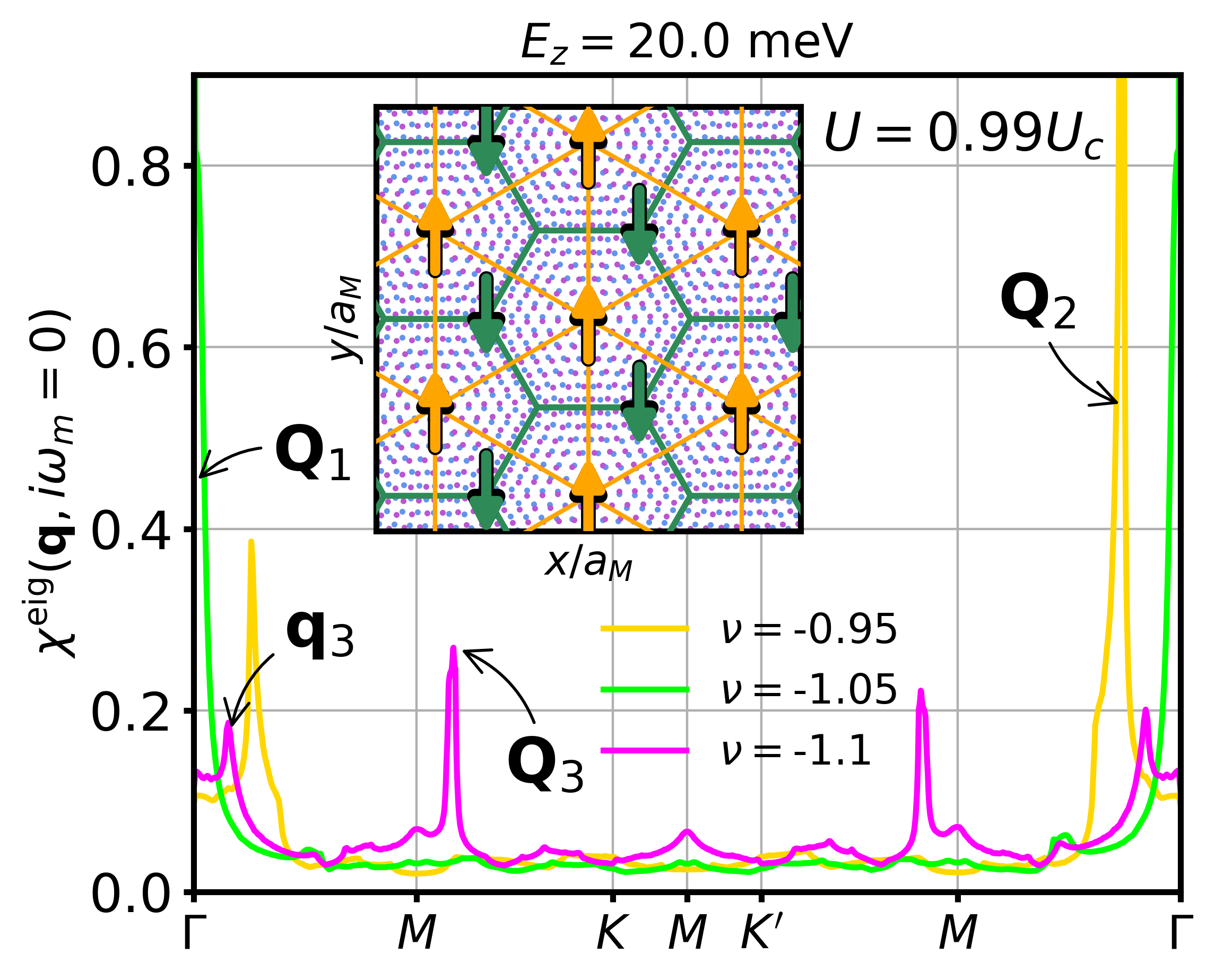}
\caption{
	Main eigenvalue of the spin-valley susceptibilities along the Brillouin zone high-symmetry directions.
    Line colors correspond to the same points in the phase diagram as the symbols in Fig.~\ref{fig:dos}(a).
    For each curve, we set the main Hubbard interaction, $U_{MX}$, to be $0.99U_c$, where $U_c$ is the critical Hubbard interaction to trigger the Stoner criterion. The inset shows the homogeneous spin-valley susceptibility matrix elements for $[{\chi}(\boldsymbol{q})]^{p\xi,p\xi}_{p\xi,p\xi}$ at the Van Hove singularity filling $\nu=-1.05$.
    }
\label{fig:chi}
\end{center}
\end{figure}
Therefore, considering the electron concentration in the $MM$ and $MX$ orbitals, our mRPA results are consistent with a commensurate spin-valley ordered phase with $MM-MM$ and $MX-MX$ ferromagnetic alignment, as well as $MM-MX$ antiferromagnetic alignment, as depicted  in the inset of Fig.~\ref{fig:chi}.
 { 
The ferromagnetic alignments resemble a valley polarized order, common in 2D materials \cite{Yi_RecentAdvancesQuantumEffects2DMaterials_2019}.
However, here the spin-valley polarization is compensated between orbitals $MM$ and $MX$, and add up to give a zero net magnetization.}

 We also note that the $XM$ orbital does not participate in the spin-valley polarization state.
This is a natural consequence of layer polarization induced by the high electric field where the magnetic instability is placed.
In fact, as shown in Fig.~\ref{fig:model}(b), the $XM-XM$ onsite energy at $E_z=20$ meV is the highest, signaling this orbital is pushed away form the Fermi level by the electric field.
Also, the $XM-MX$ hopping is further suppressed at high twist angles $\theta\sim5^o$, which effectively disconnects the two sites in the hexagonal lattice.

 This trend also seen in the Fermi surfaces of Fig.~\ref{fig:dos}(b), where we notice that the $MM$ and $MX$ orbitals are dominating the orbital contribution.
This way, they naturally better participate on the Fermi surface nesting effect associated with the Stoner instability.
This fact motivates an interpretation of Eq.~\eqref{eq:OP_result} as follows:
the order parameter exactly at the lower hole-doping side of the Van Hove singularity can be understood as an order of the type $\sigma_z\otimes\tau_z$, where $\sigma_i$ and $\tau_i$ Pauli matrices are defined in the spin space and the reduced orbital space $\{MM,MX\}$, respectively.

For filling factors slightly larger than the Van Hove value (e.g., $\nu=-0.95$, purple curve in Fig.~\ref{fig:chi}), the divergence still occurs at $\boldsymbol{Q}_2\approx0$  {and, as shown in Appendix~\ref{appendix:op}, numerically compatible with} the same type of magnetic order. 
Now, for filling factors slightly lower than the Van Hove value (e.g., $\nu=-1.1$, yellow curve in Fig.~\ref{fig:chi}), the divergence occurs for  $\boldsymbol{Q}_3\approx (\pi,0)$, near the $M$ point. In this case, the order parameter is of the form   
\begin{equation}
\begin{split}
    \Delta_+(\boldsymbol{Q} \approx \boldsymbol{Q}_3)\propto &\Delta \sum_{\boldsymbol{k}}\langle c_{(\boldsymbol{k}+\boldsymbol{Q}_3)(MM\uparrow)}^{\dagger}c_{\boldsymbol{k}(MX\downarrow)}\rangle \; ,
\end{split}
\label{eq:OP_result_2}
\end{equation}
 { In contrast to the lower hole-doping and Van Hove singularity case, further hole doping induces a strong interorbital channel of the type $\sigma_{+}\otimes\tau_+$.
This order is associated with the larger-momentum nesting vector $\boldsymbol{Q}_3$ that connects the $MM$ and $MX$ orbitals for opposite spin-valleys [see Fig.~\ref{fig:dos}(b) for $\nu=-1.1$].
}

Notice, however, that the sharp peak in $\chi^{\rm eig}(\boldsymbol{q})$ at $\boldsymbol{q}=\boldsymbol{q}_3\approx 0$ for $\nu=-1.1$ indicates that the zero-momentum contribution to spin scattering is still very strong, even for values of $U$ very close to the transition value ($U=0.99U_c$). This supports a picture in which $\boldsymbol{q}\approx0$ spin-fluctuations favor  superconductivity in this side of the Van Hove line as well. 

The effects of off-site interactions have been thoroughly discussed within the weak-coupling limit in from the functional renormalization group (fRG) perspective 
of Ref.~\cite{Fischer_Phys.Rev.X_041055_2025}.
In that work, the largest long-range Coulomb term is is estimated to be $V_{(MM)(MX)}/U_{MX}\approx0.31$.
These fRG results show that long-range Coulomb interactions do not induce relevant changes in the uncorrelated normal state Fermi surfaces, especially around the Van Hove line, where we remark that for the Stoner phase diagram, the Fermi surface is the protagonist.
Moreover, their numerical calculations also indicate that qualitative features of the phase diagram are not altered by the long-range interactions despite the appearance of another magnetic order, a spin-bond order.
The spin-bond order is non-local so that it cannot be written as a function of a nesting vector $\boldsymbol{Q}$ and, therefore, cannot be captured by our mRPA calculations.

Additionally, the charge fluctuations channel is not usually enhanced enough by long-range interactions to compete with the spin fluctuations, as previously shown in the context of magic-angle twisted bilayer graphene \cite{Braz:Phys.Rev.B:184502:2024}. Within the mRPA, the off-site interactions act by generating a Coulomb tail which makes the interaction matrix $\hat{U}$ acquire a momentum dependence $\hat{U}(\boldsymbol{q})$.
The Coulomb tail can then generate effective attractive or repulsive momentum-selective channels \cite{Braz:Phys.Rev.B:184502:2024}.
The long-range Coulomb tail in momentum space, which follows the same lattice structure as that of magic-angle twisted bilayer graphene, is repulsive and favors the Stoner instability at $\boldsymbol{Q}\sim0$ nesting vectors, so the $\boldsymbol{Q}_1$ and $\boldsymbol{Q}_2$ orders are always favored by long-range interactions \cite{Braz:Phys.Rev.B:184502:2024}.
The $\boldsymbol{Q}\approx M$ nesting found at the hole doping side of the Van Hove, however, can be sensitive to the strong off-site interactions.
This effect would replace the $\boldsymbol{Q}_3\approx \boldsymbol{M}$ by the $\boldsymbol{q}_3\approx0$ spin-valley order, which recovers the compensated spin-valley polarization order as at the Van Hove singularity.
Altogether, accounting for long-range interactions would not qualitatively affect the Stoner phase diagram presented here. 

These results can also give insights on the reentrant superconductivity observed experimentally.  In fact, while superconductivity is suppressed in the spin-valley ordered regions of the phase diagram, it tends to be highly favored in the regions located  at the paramagnetic flanks of the Stoner boundary  [green regions of Fig.~\ref{fig:stoner}(a)], especially since the peaks at $\chi^{\rm eig}(\boldsymbol{q}=0)$ for filling factors in this region will play the dominant contribution to the pairing interaction \cite{graserNeardegeneracySeveralPairing2009}. 

The strengthening of superconductivity at the edges of the Stoner boundary can be understood by taking the single-orbital Hubbard model as an example. In this case, Eq.~\eqref{eq:stoner} is algebraic and the Stoner criterion at zero temperature 
is expressed by $\alpha = \chi_0(0) U \propto \rho_0 U$, where $\rho_0$ is the density of states at the Fermi energy  
and $U$ is the Hubbard interaction strength.
On the other hand, a BCS superconductor has a critical temperature $T_c\sim e^{-1/\lambda}$ with $\lambda=\rho_0 V$ the pairing strength for a system with Fermi surface density of states $\rho_0$ and effective electron-electron interaction $V$.
At this simplified level, $U$ and $V$ are constants and we notice the similarity between $\alpha$ and $\lambda$: the density of states.
When the density of states is high enough, $\alpha>1$ and the spin-valley order sets in.
However, when $\alpha\lesssim1$, $\lambda\sim\alpha V/U$ favors the superconducting pairing.
In this way, we expect the green region in Fig.~\ref{fig:stoner}(a) to be superconducting.
These arguments have been considered quantitatively at the mean-field level \cite{LöthmanUniversalPhaseDiagramsSuperconducting2017}.

\section{Concluding remarks}
\label{sec:Conclusions}

In this work, we have investigated the interplay between spin--valley instabilities and superconductivity in twisted bilayer WSe$_2$ at a twist angle of $5^\circ$, motivated by recent experimental observations of reentrant superconducting behavior. Starting from a faithful three-orbital Wannier description of the spin--valley--locked moiré bands and incorporating orbital-dependent Hubbard interactions, we employed a matrix random phase approximation (mRPA) framework to analyze the evolution of magnetic instabilities as a function of hole filling and perpendicular electric field.

By computing the spin--valley susceptibilities across the electric-field–density phase diagram, we identified the Stoner boundary lines associated with the onset of spin--valley ordered states. These instabilities occur in close proximity to a Lifshitz transition in the topmost moiré valence band, marked by the crossing of a Van Hove singularity in the density of states. The Lifshitz transition separates regimes with distinct Fermi-surface topologies, and its presence strongly enhances spin--valley fluctuations, providing favorable conditions for interaction-driven ordering phenomena.

Our analysis reveals that the dominant spin--valley instabilities are governed by interorbital spin--valley–flip processes involving the $MM$ and $MX$ moiré orbitals. 
Importantly, the momentum structure of the leading instability changes across the Van Hove line, indicating that the spatial pattern of the spin--valley order parameter depends sensitively on carrier density.
 We remark the prediction of a compensated spin-valley polarization order triggered by the strong density of states at Van Hove singularity.
This filling-dependent momentum selection highlights the crucial role played by the multi-orbital character of the moiré bands and the underlying Lifshitz transition in shaping the nature of the ordered states.

Beyond the emergence of spin--valley order, we find that the Stoner boundary lines are accompanied by pronounced fluctuations on their flanks. 
Within the mRPA framework, these enhanced spin--valley fluctuations naturally give rise to attractive pairing interactions in nearby regions of the phase diagram. 
As a result, superconducting instabilities are usually favored in the vicinity of, but not inside, the spin--valley--ordered phases. 
This mechanism provides a microscopic explanation for a reentrant superconducting dome surrounding magnetically ordered regions, in qualitative agreement with experimental observations in twisted bilayer WSe$_2$ at $5^\circ$.

More broadly, our results underscore the central role of Fermi-surface topology and multi-orbital spin--valley physics in moiré transition-metal dichalcogenides. The close connection between a Lifshitz transition, enhanced spin--valley fluctuations, and reentrant superconductivity identified here suggests a unifying framework for understanding correlated and superconducting phases in twisted WSe$_2$ and related systems. 
Future work incorporating self-energy effects beyond mRPA, as well as a more explicit treatment of superconducting pairing channels, may further clarify the symmetry and robustness of the superconducting states and their interplay with competing orders.

\begin{acknowledgments}
We thank George Martins for insightful discussions on the mRPA method  {and Valetin Cr\'epel and Ammon Fischer for insights on modeling the weak-coupling limit to the tTMDs}. 
We thank Valentin Cr\'epel for presenting the open problem of reentrant superconductivity in $\sim5^o$ twisted bilayer WSe$_2$, and Hugo Strand and Andrew Hardy for personalized implementations of the general susceptibility within TRIQS TPRF.
We acknowledge financial support by CNPq (309789/2020-6, and 312622/2023-6), and FAPESP (grant Nos. 2022/15453-0, 2023/14902-8, and 2025/17852-7).

\section*{Data availability} Data supporting the findings of this paper are openly available on Zenodo \cite{ZenodoData_tWSe2}.

\end{acknowledgments}

\appendix
\section{Robustness of the Stoner boundary to the interaction parameters}
\label{appendix:robust}
Here we discuss the robustness of the Stoner boundary phase diagram shown in Fig.~\ref{fig:stoner} and the susceptibility nesting vectors presented in Fig.~\ref{fig:chi} to the interaction parameters used in the multiorbital Hubbard model of Eq.~\eqref{Hint}.

\begin{figure}[t]
\begin{center}
\includegraphics[width=1.0\columnwidth]{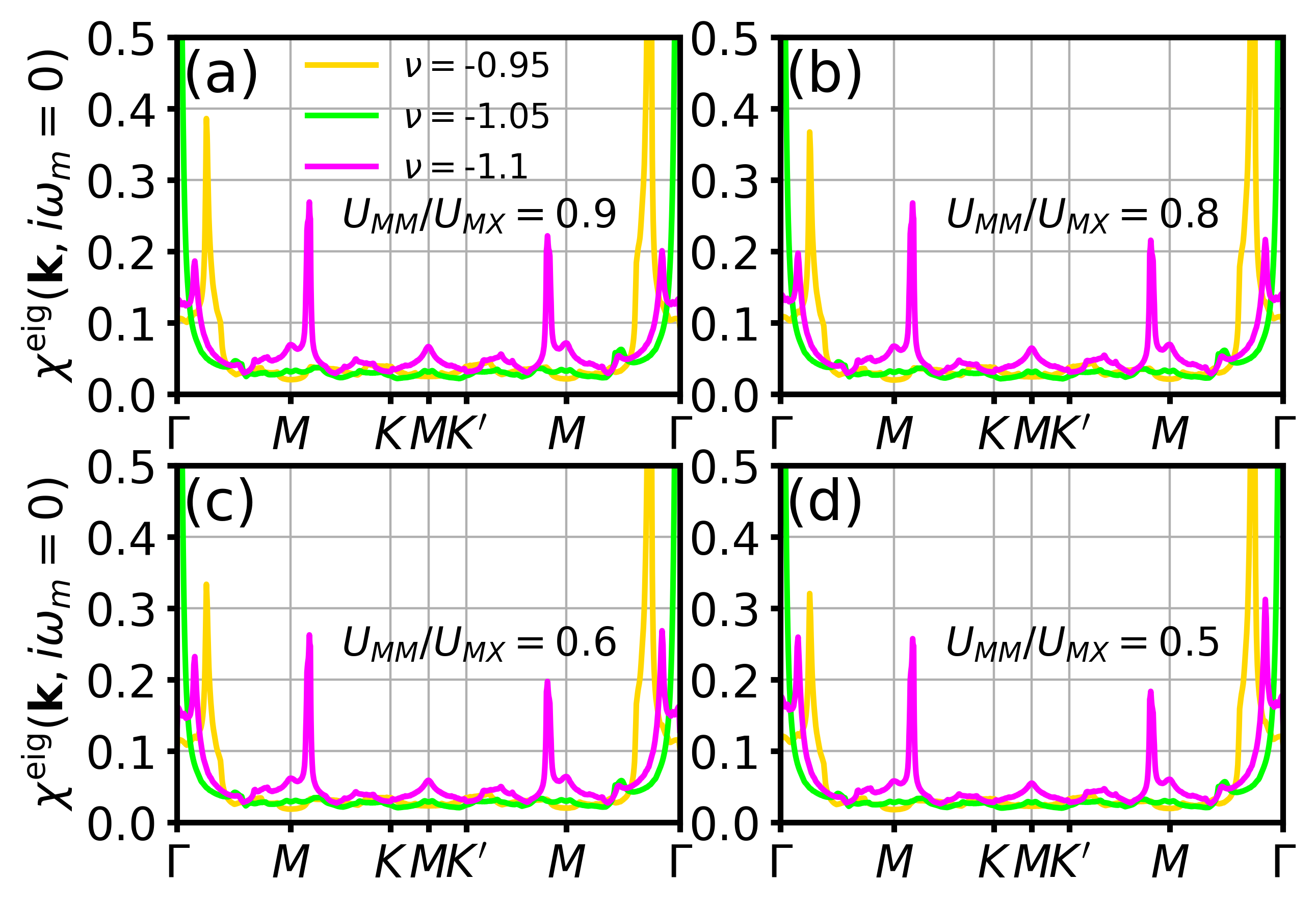}
\caption{
	Same as Fig.~\ref{fig:chi} for different ratios of $U_{MX}/U_{MM}$.
    }
\label{fig:chi_appendix}
\end{center}
\end{figure}

First, we show that the ratio $U_{MM}/U_{MX}$ is of little importance for determining the nesting vectors that give rise to the magnetic instability.
Figure \ref{fig:chi_appendix} shows the same plots as Fig.~\ref{fig:chi} for different interaction ratios.
In these figures, the Hubbard $U$ is used as a control parameter to assess the ordering vector at magnetic instability.
Explicitly, the interactions $\{U_{MM},U_{MX},U_{XM}\}$ are rewritten as $U\{U_{MM}/U_{MX},1,1\}$.
Fig.~\ref{fig:chi_appendix} shows that the fillings $\nu=-0.95$ and $\nu=-1.05$ are fully robust to $U_{MM}/U_{MX}$ changes.
For $\nu=-1.1$, as shown in Fig.~\ref{fig:chi_appendix}(b), a decrease of 10$\%$ in the $U_{MM}/U_{MX}$ ratio still keeps the $\boldsymbol{Q}_3$ nesting vector.
On the other hand, unrealistic further imbalance between the hexagonal $\{MX,XM\}$ and the triangular $\{MM\}$ lattices tends to slightly favor a $\boldsymbol{q}\approx0$ nesting, more similarly to the lower hole-doping cases.
The order parameter in such cases is therefore the one of Eq.~\eqref{eq:OP_result}.

Having established that the unbalance between the hexagonal and triangular lattices is of little influence to the magnetic phases reported in the main text, we now show in Fig.~\ref{fig:stoner_appendix} the Stoner phase diagram of Fig.~\ref{fig:stoner} for different values of $U_{MX}$, while keeping $U_{MM}/U_{MX}=0.9$.
Figure \ref{fig:stoner_appendix}(a) shows that a suppression of 5$\%$ deviates the emergence of the superconducting phase from the $\nu=-1$ line, while enhancing 5, 10 and 15$\%$ of the interaction pushes the ordered states to further lower hole-doping ranges.
As also argued in the end of Sec.~\ref{sec:spin_valley}, this trend is a natural consequence of the Stoner criterion, which depends on a balance of the density of states and the interaction strength.
Higher $U$ then triggers the Stoner instability for lower $\rho_0$.
However, we remark that the Stoner boundaries shown in green are still robust to changes on the onset of the magnetic instability in the phase diagram.

\begin{figure}[t]
\begin{center}
\includegraphics[width=1.0\columnwidth]{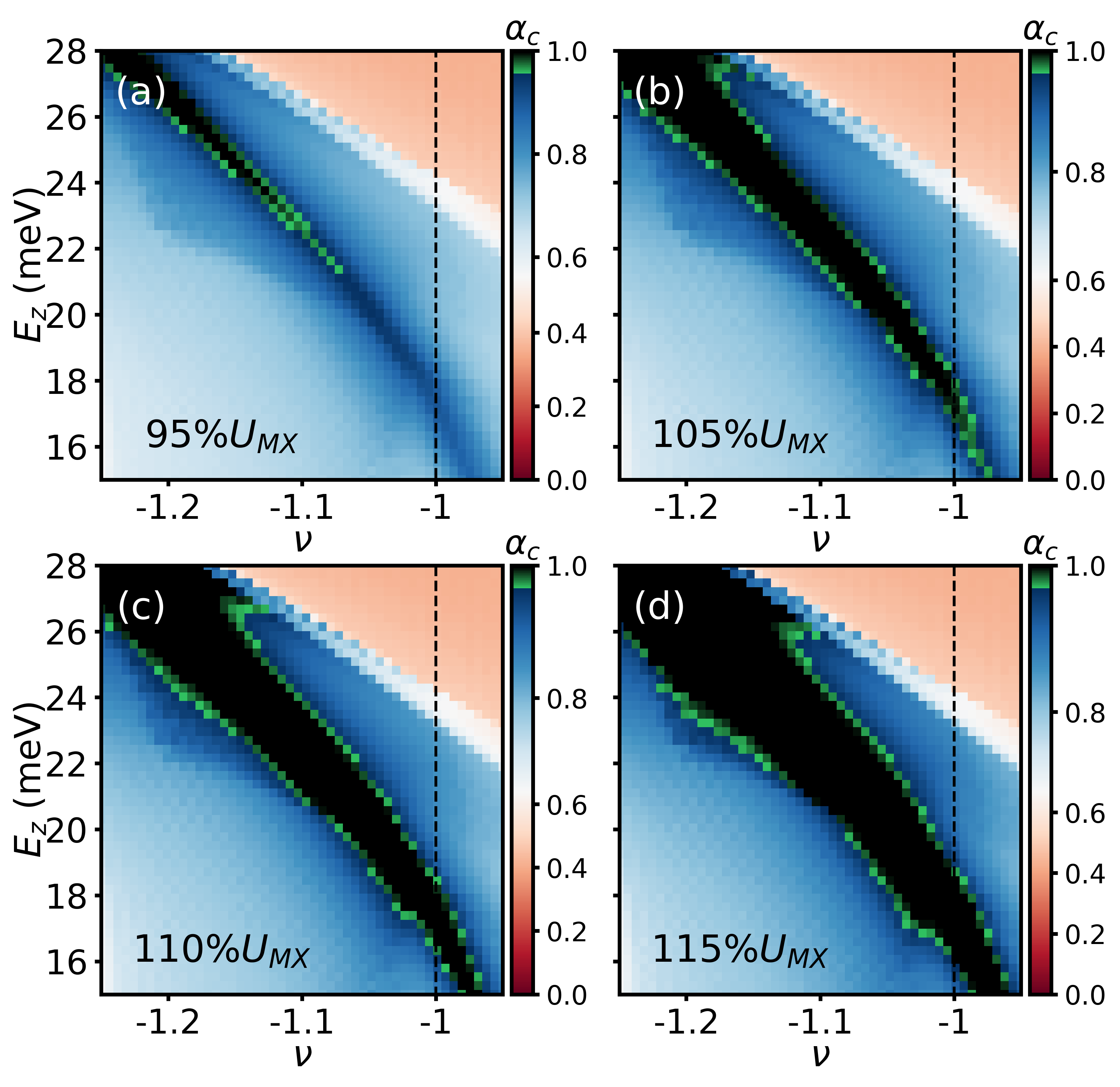}
\caption{
	Same as Fig.~\ref{fig:stoner} for different $U_{MX}$ as compared with the main text value $U_{MX}=41.3$ meV.
    }
\label{fig:stoner_appendix}
\end{center}
\end{figure}
%

\section{Efficient computation of the spin-valley susceptibility}
\label{appendix:efficiency}
Computing the susceptibility of a multiorbital system is a formidable problem.
The susceptibility is defined in the two-body basis formed by one electron that can tunnel from spin-valley orbital $t\xi_4$ to $p\xi_1$ which scatters with another electron tunneling from $q\xi_2$ to $r\xi_3$.

With $N_\xi$ spin-valley indices and $N_o$ orbitals, this basis has dimension $N_\xi^2N_o^2$, which in the susceptibility translates to a matrix with $N_\xi^4N_o^4$ elements.
For each of these elements, the momentum and imaginary frequency summations of Eq.~\eqref{eq:chi_0td} must be performed.
With $N_{\boldsymbol{k}}$ $\boldsymbol{k}$ points and $N_\omega$ imaginary frequencies, using Fourier transformations makes the problem simpler, and computations grow with $N_\xi^4N_o^4N_{\boldsymbol{k}}N_{\omega}\ln (N_{\boldsymbol{k}}N_{\omega})$ complexity \cite{KayeDiscreteLehmannRepresentationImaginary2022} as implemented in the TPRF package \cite{StrandTPRFToolboxTRIQS2019} of the TRIQS library \cite{ParcolletTRIQSToolboxForResearch2015}.
This is in contrast with the analytical expression for the susceptibility which requires $N_\xi^4N_o^4N_{\boldsymbol{k}}^2$ operations \cite{graserNeardegeneracySeveralPairing2009}.
Here, $N_{o}=3$ and $N_\xi=2$, so, in principle, 1296 summations should be performed.
However, there is one aspect of the non-interacting model that helps in this task: The two spin-valleys are decoupled in the non-interacing Hamiltonian $H_0$.
With this in mind, we note that the only non-zero matrix elements of the bare susceptibility are those such that $\xi_1=\xi_4$ and $\xi_2=\xi_3$ for which the Green's functions of Eq.~\eqref{eq:chi_0td} keep only intra-spin-valley terms.
Omitting the orbital indices for conciseness, we only need to compute the susceptibility elements $\left[\chi_{0}(\boldsymbol{q},0)\right]_{\uparrow , \uparrow}^{\uparrow, \uparrow}$, $\left[\chi_{0}(\boldsymbol{q},0)\right]_{\downarrow , \downarrow}^{\downarrow, \downarrow}$, $\left[\chi_{0}(\boldsymbol{q},0)\right]_{\uparrow , \downarrow}^{\downarrow, \uparrow}$, and $\left[\chi_{0}(\boldsymbol{q},0)\right]_{\downarrow , \uparrow}^{\uparrow, \downarrow}$ in the three-orbital space ($N_o=3$).
Thus, instead of one $N_oN_\xi=6$ problem, we end up with four $N_o=3$ problems, which lowers the number of matrix elements computed from 1296 to 324.

\section{Magnetic order parameters at the instability}
\label{appendix:op}
In this Appendix, we detail the numerical procedure to obtain the magnetic order parameter Eq.~\eqref{eq:OP} at the instability from the particle-particle bubble Eq.~\eqref{eq:chi}.

The order parameter probed by the mRPA particle-hole susceptibility can only give information at the magnetic instability.
To that, any analysis must be done by first placing the system close to the instability.
Here, this is done by changing the control parameter $U$ as defined in Appendix \ref{appendix:robust} to ratios close to the critical parameter $U_c$.
For this reason, we show Fig.~\ref{fig:chi} for $U=0.99U_c$.
Then, we select the nesting vector, which is the momentum vector with the larger susceptibility eigenvalue.
Fig.~\ref{fig:chi} shows the respective vector positions as $\boldsymbol{Q}_1$, $\boldsymbol{Q}_2$ and $\boldsymbol{Q}_3$ for fillings $\nu=-0.95$, $-1.05$, and $-1.1$, respectively.

Figure~\ref{fig:op} shows the eigenvectors respective to the maximum eigenvalues at the instability in the multiorbital basis. 
We show the real part (black), imaginary part (red), and absolute value (blue) curves.
Figures~\ref{fig:op}(a) and (b) show results for $\boldsymbol{Q}_1$ and $\boldsymbol{Q}_2$, respectively, but which share roughly the same orbital structure, revealing a similarity for the $\boldsymbol{q}\approx0$ orders.
In Eq.~\eqref{eq:OP_result}, we report the two main contributions for the order parameter components, which are place at $p\xi t\xi'=MM\uparrow MM\uparrow$ and $p\xi t\xi'=MX\downarrow MX\downarrow$.
In both cases, $\nu=-0.95$ and $\nu=-1.05$, these two components have opposite signs and the same amplitude, which is equivalent to the $\sigma_z\otimes\tau_z$ order reported in the main text.
When comparing $\nu=-0.95$ and $\nu=-1.05$, one may notice that the eigenvector components differ by a global minus sign.
We chose to leave this numerical artifact untouched to exemplify that the diagonalization procedure has an angular gauge that can, for example, flip sign in the eigenvectors.

\begin{figure}[t]
\begin{center}
\includegraphics[width=1.0\columnwidth]{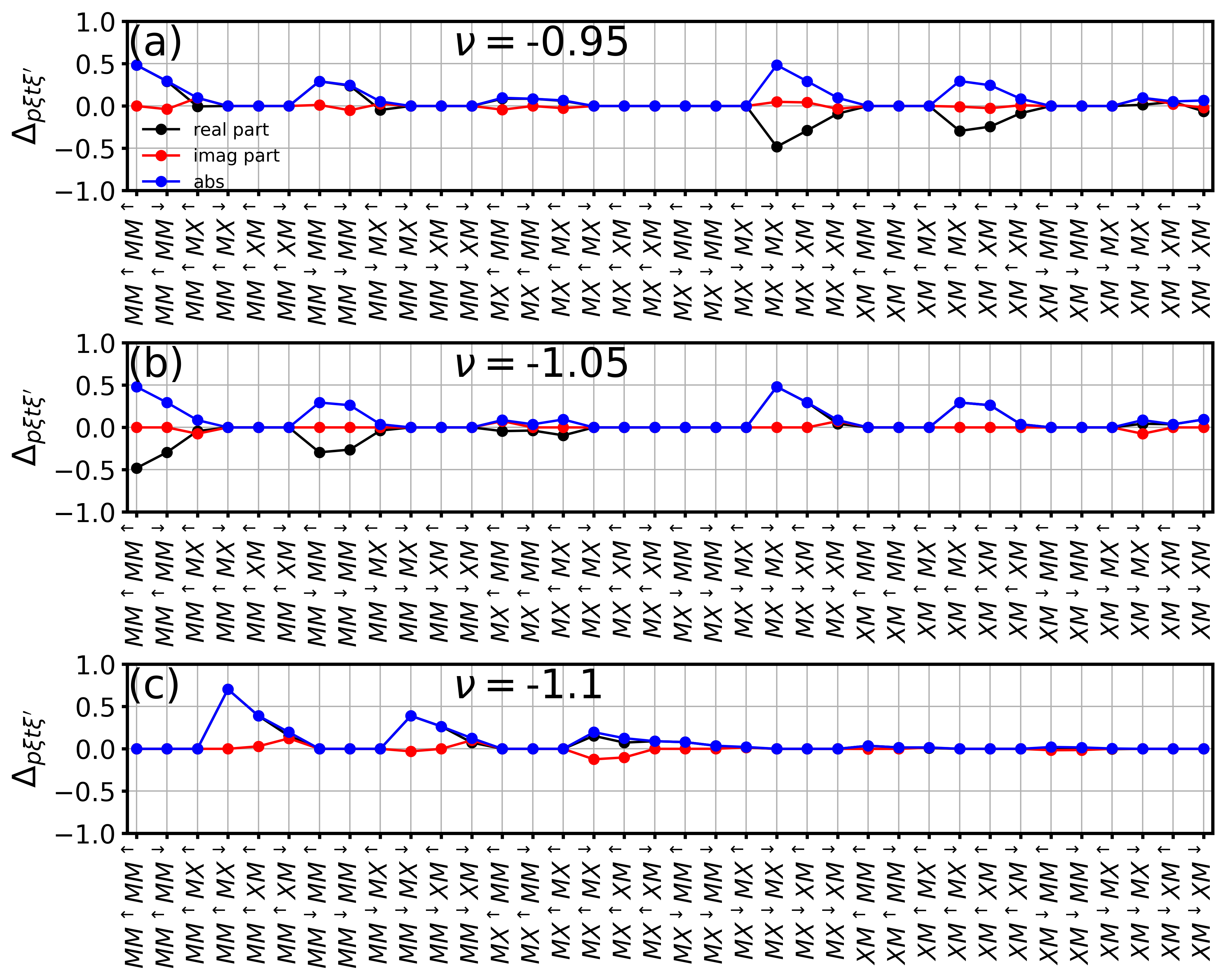}
\caption{
	Elements of the spin-valley order parameter at the instability for fillings $\nu=-0.95$ (a), $\nu=-1.05$ (b), and $\nu=-1.1$ (c) at $E_z=20$ meV. The real part (black), imaginary part (red), and absolute value (blue) are shown.
    }
\label{fig:op}
\end{center}
\end{figure}

Figure~\ref{fig:op}(c) shows the order parameter components for the $\nu=-1.1$ filling at the $\boldsymbol{Q}_3$ nesting vector.
The dominating element of the ordering vector is the single $p\xi t\xi'=MM\uparrow MX\downarrow$, which determines the $\sigma_+\otimes\tau_+$ component shown in Eq.~\eqref{eq:OP_result_2}.
We remark that, following the analysis of Appendix~\ref{appendix:robust}, if the unbalance between hexagonal and triangular lattice interaction strength is too large, then the $\boldsymbol{Q}_3$ nesting is replaced by a $\boldsymbol{Q}\approx0$ order, which should be in agreement with the compensated spin-valley polarization order found for $\nu=-0.95$ and $\nu=-1.05$.

Finally, to further support the findings highlighted by the order parameter at the instability, we cross-check those predictions with the susceptibility matrix elements close to the instability.
Fig.~\ref{fig:chi_elements} shows such matrix elements for all the three example fillings studied.
We note the instability at the same scattering vectors as shown in Fig.~\ref{fig:chi} for the matrix elements pointed out by the order parameters in Fig.~\ref{fig:op}.
We remark that the divergences occur at homogeneous matrix elements $[{\chi}(\boldsymbol{q})]^{p\xi,p'\xi'}_{p\xi,p'\xi'}$ for $\nu=-0.95$ [Fig.~\ref{fig:op}(a)] and $\nu=-1.05$ [Fig.~\ref{fig:op}(b)], whereas the off-diagonal channel $[{\chi}(\boldsymbol{q})]^{p\xi,p'\xi'}_{p'\xi',p\xi}$ is leading at $\nu=-1.1$ [Fig.~\ref{fig:op}(c)].

\begin{figure}[t]
\begin{center}
\includegraphics[width=1.0\columnwidth]{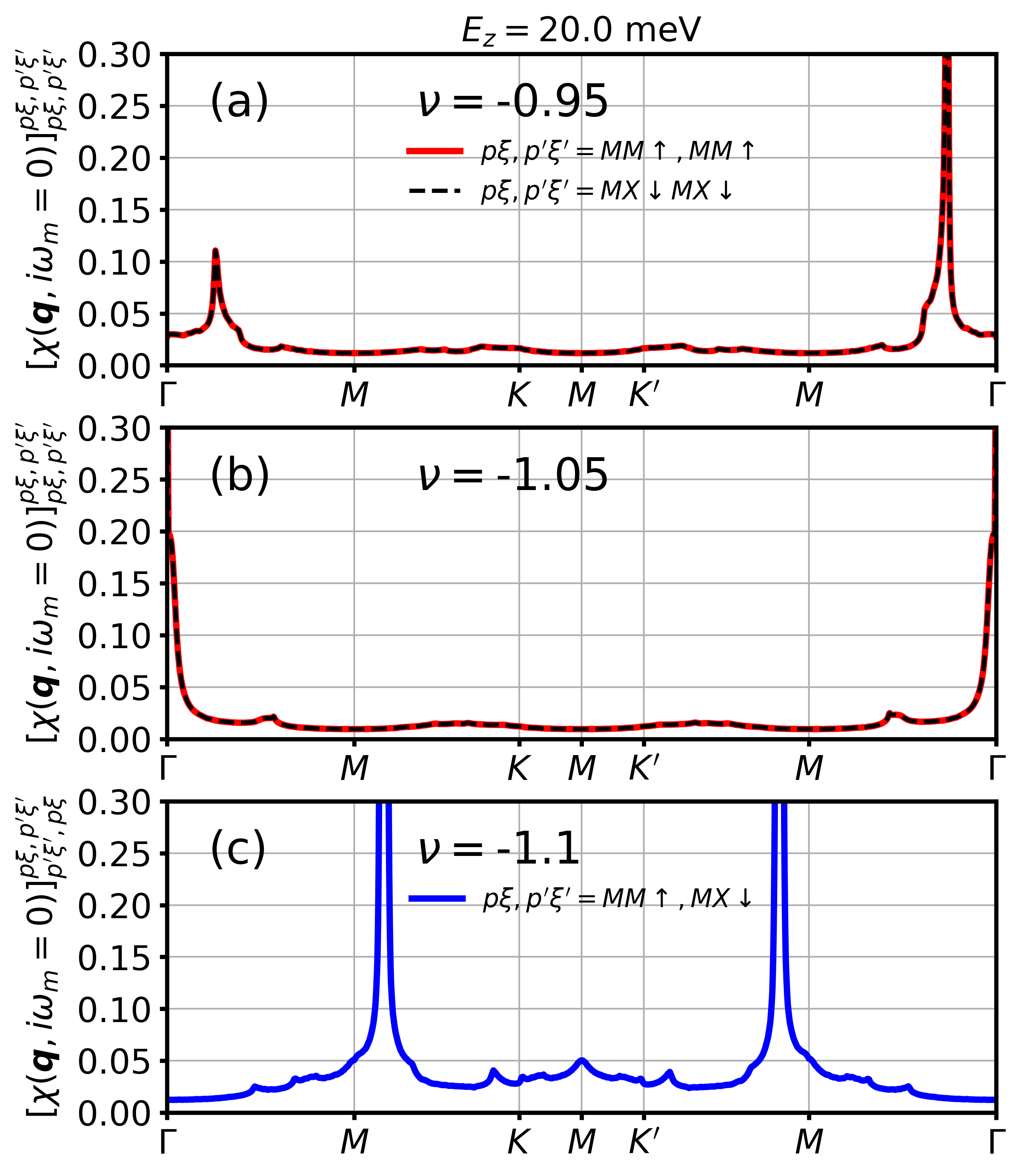}
\caption{
	Larger bare spin-valley susceptibility matrix elements for $[{\chi}(\boldsymbol{q})]^{p\xi,p'\xi'}_{p\xi,p'\xi'}$ at the fillings $\nu=-0.95$ (a) and $\nu=-1.05$ (b), and $[{\chi}(\boldsymbol{q})]^{p\xi,p'\xi'}_{p'\xi',p\xi}$ at $\nu=-1.1$ (c).
    We adopt $U=0.99U_c$ in all cases.
    }
\label{fig:chi_elements}
\end{center}
\end{figure}
%


\begin{thebibliography}{56}%
\makeatletter
\providecommand \@ifxundefined [1]{%
 \@ifx{#1\undefined}
}%
\providecommand \@ifnum [1]{%
 \ifnum #1\expandafter \@firstoftwo
 \else \expandafter \@secondoftwo
 \fi
}%
\providecommand \@ifx [1]{%
 \ifx #1\expandafter \@firstoftwo
 \else \expandafter \@secondoftwo
 \fi
}%
\providecommand \natexlab [1]{#1}%
\providecommand \enquote  [1]{``#1''}%
\providecommand \bibnamefont  [1]{#1}%
\providecommand \bibfnamefont [1]{#1}%
\providecommand \citenamefont [1]{#1}%
\providecommand \href@noop [0]{\@secondoftwo}%
\providecommand \href [0]{\begingroup \@sanitize@url \@href}%
\providecommand \@href[1]{\@@startlink{#1}\@@href}%
\providecommand \@@href[1]{\endgroup#1\@@endlink}%
\providecommand \@sanitize@url [0]{\catcode `\\12\catcode `\$12\catcode
  `\&12\catcode `\#12\catcode `\^12\catcode `\_12\catcode `\%12\relax}%
\providecommand \@@startlink[1]{}%
\providecommand \@@endlink[0]{}%
\providecommand \url  [0]{\begingroup\@sanitize@url \@url }%
\providecommand \@url [1]{\endgroup\@href {#1}{\urlprefix }}%
\providecommand \urlprefix  [0]{URL }%
\providecommand \Eprint [0]{\href }%
\providecommand \doibase [0]{https://doi.org/}%
\providecommand \selectlanguage [0]{\@gobble}%
\providecommand \bibinfo  [0]{\@secondoftwo}%
\providecommand \bibfield  [0]{\@secondoftwo}%
\providecommand \translation [1]{[#1]}%
\providecommand \BibitemOpen [0]{}%
\providecommand \bibitemStop [0]{}%
\providecommand \bibitemNoStop [0]{.\EOS\space}%
\providecommand \EOS [0]{\spacefactor3000\relax}%
\providecommand \BibitemShut  [1]{\csname bibitem#1\endcsname}%
\let\auto@bib@innerbib\@empty
\bibitem [{\citenamefont {Andrei}\ \emph {et~al.}(2021)\citenamefont {Andrei},
  \citenamefont {Efetov}, \citenamefont {Jarillo-Herrero}, \citenamefont
  {MacDonald}, \citenamefont {Mak}, \citenamefont {Senthil}, \citenamefont
  {Tutuc}, \citenamefont {Yazdani},\ and\ \citenamefont
  {Young}}]{AndreiMarvelsOfMoiréMaterials2021}%
  \BibitemOpen
  \bibfield  {author} {\bibinfo {author} {\bibfnamefont {E.~Y.}\ \bibnamefont
  {Andrei}}, \bibinfo {author} {\bibfnamefont {D.~K.}\ \bibnamefont {Efetov}},
  \bibinfo {author} {\bibfnamefont {P.}~\bibnamefont {Jarillo-Herrero}},
  \bibinfo {author} {\bibfnamefont {A.~H.}\ \bibnamefont {MacDonald}}, \bibinfo
  {author} {\bibfnamefont {K.~F.}\ \bibnamefont {Mak}}, \bibinfo {author}
  {\bibfnamefont {T.}~\bibnamefont {Senthil}}, \bibinfo {author} {\bibfnamefont
  {E.}~\bibnamefont {Tutuc}}, \bibinfo {author} {\bibfnamefont
  {A.}~\bibnamefont {Yazdani}},\ and\ \bibinfo {author} {\bibfnamefont {A.~F.}\
  \bibnamefont {Young}},\ }\bibfield  {title} {\bibinfo {title} {The marvels of
  moiré materials},\ }\href {https://doi.org/10.1038/s41578-021-00284-1}
  {\bibfield  {journal} {\bibinfo  {journal} {Nat. Rev. Mat.}\ }\textbf
  {\bibinfo {volume} {6}},\ \bibinfo {pages} {201} (\bibinfo {year}
  {2021})}\BibitemShut {NoStop}%
\bibitem [{\citenamefont {Cao}\ \emph {et~al.}()\citenamefont {Cao},
  \citenamefont {Fatemi}, \citenamefont {Fang}, \citenamefont {Watanabe},
  \citenamefont {Taniguchi}, \citenamefont {Kaxiras},\ and\ \citenamefont
  {Jarillo-Herrero}}]{cao_unconventional_2018}%
  \BibitemOpen
  \bibfield  {author} {\bibinfo {author} {\bibfnamefont {Y.}~\bibnamefont
  {Cao}}, \bibinfo {author} {\bibfnamefont {V.}~\bibnamefont {Fatemi}},
  \bibinfo {author} {\bibfnamefont {S.}~\bibnamefont {Fang}}, \bibinfo {author}
  {\bibfnamefont {K.}~\bibnamefont {Watanabe}}, \bibinfo {author}
  {\bibfnamefont {T.}~\bibnamefont {Taniguchi}}, \bibinfo {author}
  {\bibfnamefont {E.}~\bibnamefont {Kaxiras}},\ and\ \bibinfo {author}
  {\bibfnamefont {P.}~\bibnamefont {Jarillo-Herrero}},\ }\bibfield  {title}
  {\bibinfo {title} {Unconventional superconductivity in magic-angle graphene
  superlattices},\ }\href {https://doi.org/10.1038/nature26160} {\bibfield
  {journal} {\bibinfo  {journal} {Nature}\ }\textbf {\bibinfo {volume} {556}},\
  \bibinfo {pages} {43}}\BibitemShut {NoStop}%
\bibitem [{\citenamefont {Li}\ \emph {et~al.}(2026)\citenamefont {Li},
  \citenamefont {Qiu}, \citenamefont {Wu},\ and\ \citenamefont
  {MacDonald}}]{Li_NatScienceRev_2026}%
  \BibitemOpen
  \bibfield  {author} {\bibinfo {author} {\bibfnamefont {B.}~\bibnamefont
  {Li}}, \bibinfo {author} {\bibfnamefont {W.-X.}\ \bibnamefont {Qiu}},
  \bibinfo {author} {\bibfnamefont {F.}~\bibnamefont {Wu}},\ and\ \bibinfo
  {author} {\bibfnamefont {A.~H.}\ \bibnamefont {MacDonald}},\ }\bibfield
  {title} {\bibinfo {title} {Quantum phases in twisted homobilayer transition
  metal dichalcogenides},\ }\href {https://doi.org/10.1093/nsr/nwaf570}
  {\bibfield  {journal} {\bibinfo  {journal} {National Science Review}\
  }\textbf {\bibinfo {volume} {13}},\ \bibinfo {pages} {nwaf570} (\bibinfo
  {year} {2026})}\BibitemShut {NoStop}%
\bibitem [{\citenamefont {Wang}\ \emph {et~al.}(2020)\citenamefont {Wang},
  \citenamefont {Shih}, \citenamefont {Ghiotto}, \citenamefont {Xian},
  \citenamefont {Rhodes}, \citenamefont {Tan}, \citenamefont {Claassen},
  \citenamefont {Kennes}, \citenamefont {Kim}, \citenamefont {Watanabe},
  \citenamefont {Taniguchi}, \citenamefont {Zhu}, \citenamefont {Hone},
  \citenamefont {Rubio}, \citenamefont {Pasupathy},\ and\ \citenamefont
  {Dean}}]{wangCorrelatedElectronic2020}%
  \BibitemOpen
  \bibfield  {author} {\bibinfo {author} {\bibfnamefont {L.}~\bibnamefont
  {Wang}}, \bibinfo {author} {\bibfnamefont {E.-M.}\ \bibnamefont {Shih}},
  \bibinfo {author} {\bibfnamefont {A.}~\bibnamefont {Ghiotto}}, \bibinfo
  {author} {\bibfnamefont {L.}~\bibnamefont {Xian}}, \bibinfo {author}
  {\bibfnamefont {D.~A.}\ \bibnamefont {Rhodes}}, \bibinfo {author}
  {\bibfnamefont {C.}~\bibnamefont {Tan}}, \bibinfo {author} {\bibfnamefont
  {M.}~\bibnamefont {Claassen}}, \bibinfo {author} {\bibfnamefont
  {Y.}~\bibnamefont {Kennes}, \bibfnamefont {Dante M.and~Bai}}, \bibinfo
  {author} {\bibfnamefont {B.}~\bibnamefont {Kim}}, \bibinfo {author}
  {\bibfnamefont {K.}~\bibnamefont {Watanabe}}, \bibinfo {author}
  {\bibfnamefont {T.}~\bibnamefont {Taniguchi}}, \bibinfo {author}
  {\bibfnamefont {X.}~\bibnamefont {Zhu}}, \bibinfo {author} {\bibfnamefont
  {J.}~\bibnamefont {Hone}}, \bibinfo {author} {\bibfnamefont {A.}~\bibnamefont
  {Rubio}}, \bibinfo {author} {\bibfnamefont {A.~N.}\ \bibnamefont
  {Pasupathy}},\ and\ \bibinfo {author} {\bibfnamefont {C.~R.}\ \bibnamefont
  {Dean}},\ }\bibfield  {title} {\bibinfo {title} {Correlated electronic phases
  in twisted bilayer transition metal dichalcogenides},\ }\href
  {https://doi.org/10.1038/s41563-020-0708-6} {\bibfield  {journal} {\bibinfo
  {journal} {Nat. Mat.}\ }\textbf {\bibinfo {volume} {19}},\ \bibinfo {pages}
  {861} (\bibinfo {year} {2020})}\BibitemShut {NoStop}%
\bibitem [{\citenamefont {Xu}\ \emph {et~al.}(2022)\citenamefont {Xu},
  \citenamefont {Kang}, \citenamefont {Watanabe}, \citenamefont {Taniguchi},
  \citenamefont {Mak},\ and\ \citenamefont
  {Shan}}]{xuTunableBilayerHubbard2022}%
  \BibitemOpen
  \bibfield  {author} {\bibinfo {author} {\bibfnamefont {Y.}~\bibnamefont
  {Xu}}, \bibinfo {author} {\bibfnamefont {K.}~\bibnamefont {Kang}}, \bibinfo
  {author} {\bibfnamefont {K.}~\bibnamefont {Watanabe}}, \bibinfo {author}
  {\bibfnamefont {T.}~\bibnamefont {Taniguchi}}, \bibinfo {author}
  {\bibfnamefont {K.~F.}\ \bibnamefont {Mak}},\ and\ \bibinfo {author}
  {\bibfnamefont {J.}~\bibnamefont {Shan}},\ }\bibfield  {title} {\bibinfo
  {title} {A tunable bilayer hubbard model in twisted {{WSe$_2$}}},\ }\href
  {https://doi.org/10.1038/s41565-022-01180-7} {\bibfield  {journal} {\bibinfo
  {journal} {Nat. Nanotechnol.}\ }\textbf {\bibinfo {volume} {17}},\ \bibinfo
  {pages} {934} (\bibinfo {year} {2022})}\BibitemShut {NoStop}%
\bibitem [{\citenamefont {Anderson}\ \emph {et~al.}(2023)\citenamefont
  {Anderson}, \citenamefont {Fan}, \citenamefont {Cai}, \citenamefont
  {Holtzmann}, \citenamefont {Taniguchi}, \citenamefont {Watanabe},
  \citenamefont {Xiao}, \citenamefont {Yao},\ and\ \citenamefont
  {Xu}}]{andersonCorrelatedMagnetic2023}%
  \BibitemOpen
  \bibfield  {author} {\bibinfo {author} {\bibfnamefont {E.}~\bibnamefont
  {Anderson}}, \bibinfo {author} {\bibfnamefont {F.-R.}\ \bibnamefont {Fan}},
  \bibinfo {author} {\bibfnamefont {J.}~\bibnamefont {Cai}}, \bibinfo {author}
  {\bibfnamefont {W.}~\bibnamefont {Holtzmann}}, \bibinfo {author}
  {\bibfnamefont {T.}~\bibnamefont {Taniguchi}}, \bibinfo {author}
  {\bibfnamefont {K.}~\bibnamefont {Watanabe}}, \bibinfo {author}
  {\bibfnamefont {D.}~\bibnamefont {Xiao}}, \bibinfo {author} {\bibfnamefont
  {W.}~\bibnamefont {Yao}},\ and\ \bibinfo {author} {\bibfnamefont
  {X.}~\bibnamefont {Xu}},\ }\bibfield  {title} {\bibinfo {title} {Programming
  correlated magnetic states with gate-controlled moiré geometry},\ }\href
  {https://doi.org/10.1126/science.adg4268} {\bibfield  {journal} {\bibinfo
  {journal} {Science}\ }\textbf {\bibinfo {volume} {381}},\ \bibinfo {pages}
  {325} (\bibinfo {year} {2023})}\BibitemShut {NoStop}%
\bibitem [{\citenamefont {Cai}\ \emph {et~al.}(2023)\citenamefont {Cai},
  \citenamefont {Anderson}, \citenamefont {Wang}, \citenamefont {Zhang},
  \citenamefont {Liu}, \citenamefont {Holtzmann}, \citenamefont {Zhang},
  \citenamefont {Fan}, \citenamefont {Taniguchi}, \citenamefont {Watanabe},
  \citenamefont {Ran}, \citenamefont {Cao}, \citenamefont {Fu}, \citenamefont
  {Xiao}, \citenamefont {Yao},\ and\ \citenamefont
  {Xu}}]{caiFractionalQuantumHall2023}%
  \BibitemOpen
  \bibfield  {author} {\bibinfo {author} {\bibfnamefont {J.}~\bibnamefont
  {Cai}}, \bibinfo {author} {\bibfnamefont {E.}~\bibnamefont {Anderson}},
  \bibinfo {author} {\bibfnamefont {C.}~\bibnamefont {Wang}}, \bibinfo {author}
  {\bibfnamefont {X.}~\bibnamefont {Zhang}}, \bibinfo {author} {\bibfnamefont
  {X.}~\bibnamefont {Liu}}, \bibinfo {author} {\bibfnamefont {W.}~\bibnamefont
  {Holtzmann}}, \bibinfo {author} {\bibfnamefont {Y.}~\bibnamefont {Zhang}},
  \bibinfo {author} {\bibfnamefont {F.}~\bibnamefont {Fan}}, \bibinfo {author}
  {\bibfnamefont {T.}~\bibnamefont {Taniguchi}}, \bibinfo {author}
  {\bibfnamefont {K.}~\bibnamefont {Watanabe}}, \bibinfo {author}
  {\bibfnamefont {Y.}~\bibnamefont {Ran}}, \bibinfo {author} {\bibfnamefont
  {T.}~\bibnamefont {Cao}}, \bibinfo {author} {\bibfnamefont {L.}~\bibnamefont
  {Fu}}, \bibinfo {author} {\bibfnamefont {D.}~\bibnamefont {Xiao}}, \bibinfo
  {author} {\bibfnamefont {W.}~\bibnamefont {Yao}},\ and\ \bibinfo {author}
  {\bibfnamefont {X.}~\bibnamefont {Xu}},\ }\bibfield  {title} {\bibinfo
  {title} {Signatures of fractional quantum anomalous hall states in twisted
  {{MoTe$_2$}}},\ }\href {https://doi.org/10.1038/s41586-023-06289-w}
  {\bibfield  {journal} {\bibinfo  {journal} {Nature}\ }\textbf {\bibinfo
  {volume} {622}},\ \bibinfo {pages} {63} (\bibinfo {year} {2023})}\BibitemShut
  {NoStop}%
\bibitem [{\citenamefont {Zeng}\ \emph {et~al.}(2023)\citenamefont {Zeng},
  \citenamefont {Xia}, \citenamefont {Kang}, \citenamefont {Zhu}, \citenamefont
  {Knüppel}, \citenamefont {Vaswani}, \citenamefont {Watanabe}, \citenamefont
  {Taniguchi}, \citenamefont {Mak},\ and\ \citenamefont
  {Shan}}]{zengFractionalChernInsulator2023}%
  \BibitemOpen
  \bibfield  {author} {\bibinfo {author} {\bibfnamefont {Y.}~\bibnamefont
  {Zeng}}, \bibinfo {author} {\bibfnamefont {Z.}~\bibnamefont {Xia}}, \bibinfo
  {author} {\bibfnamefont {K.}~\bibnamefont {Kang}}, \bibinfo {author}
  {\bibfnamefont {J.}~\bibnamefont {Zhu}}, \bibinfo {author} {\bibfnamefont
  {P.}~\bibnamefont {Knüppel}}, \bibinfo {author} {\bibfnamefont
  {C.}~\bibnamefont {Vaswani}}, \bibinfo {author} {\bibfnamefont
  {K.}~\bibnamefont {Watanabe}}, \bibinfo {author} {\bibfnamefont
  {T.}~\bibnamefont {Taniguchi}}, \bibinfo {author} {\bibfnamefont {K.~F.}\
  \bibnamefont {Mak}},\ and\ \bibinfo {author} {\bibfnamefont {J.}~\bibnamefont
  {Shan}},\ }\bibfield  {title} {\bibinfo {title} {Thermodynamic evidence of
  fractional chern insulator in moiré {{MoTe$_2$}}},\ }\href
  {https://doi.org/10.1038/s41586-023-06452-3} {\bibfield  {journal} {\bibinfo
  {journal} {Nature}\ }\textbf {\bibinfo {volume} {622}},\ \bibinfo {pages}
  {69} (\bibinfo {year} {2023})}\BibitemShut {NoStop}%
\bibitem [{\citenamefont {Park}\ \emph {et~al.}(2023)\citenamefont {Park},
  \citenamefont {Cai}, \citenamefont {Anderson}, \citenamefont {Zhang},
  \citenamefont {Zhu}, \citenamefont {Liu}, \citenamefont {Wang}, \citenamefont
  {Holtzmann}, \citenamefont {Hu}, \citenamefont {Liu}, \citenamefont
  {Taniguchi}, \citenamefont {Watanabe}, \citenamefont {Chu}, \citenamefont
  {Cao}, \citenamefont {Fu}, \citenamefont {Yao}, \citenamefont {Chang},
  \citenamefont {Cobden}, \citenamefont {Xiao},\ and\ \citenamefont
  {Xu}}]{parkFractionalyQuantized2023}%
  \BibitemOpen
  \bibfield  {author} {\bibinfo {author} {\bibfnamefont {H.}~\bibnamefont
  {Park}}, \bibinfo {author} {\bibfnamefont {J.}~\bibnamefont {Cai}}, \bibinfo
  {author} {\bibfnamefont {E.}~\bibnamefont {Anderson}}, \bibinfo {author}
  {\bibfnamefont {Y.}~\bibnamefont {Zhang}}, \bibinfo {author} {\bibfnamefont
  {J.}~\bibnamefont {Zhu}}, \bibinfo {author} {\bibfnamefont {X.}~\bibnamefont
  {Liu}}, \bibinfo {author} {\bibfnamefont {C.}~\bibnamefont {Wang}}, \bibinfo
  {author} {\bibfnamefont {W.}~\bibnamefont {Holtzmann}}, \bibinfo {author}
  {\bibfnamefont {C.}~\bibnamefont {Hu}}, \bibinfo {author} {\bibfnamefont
  {Z.}~\bibnamefont {Liu}}, \bibinfo {author} {\bibfnamefont {T.}~\bibnamefont
  {Taniguchi}}, \bibinfo {author} {\bibfnamefont {K.}~\bibnamefont {Watanabe}},
  \bibinfo {author} {\bibfnamefont {J.-H.}\ \bibnamefont {Chu}}, \bibinfo
  {author} {\bibfnamefont {T.}~\bibnamefont {Cao}}, \bibinfo {author}
  {\bibfnamefont {L.}~\bibnamefont {Fu}}, \bibinfo {author} {\bibfnamefont
  {W.}~\bibnamefont {Yao}}, \bibinfo {author} {\bibfnamefont {C.-Z.}\
  \bibnamefont {Chang}}, \bibinfo {author} {\bibfnamefont {D.}~\bibnamefont
  {Cobden}}, \bibinfo {author} {\bibfnamefont {D.}~\bibnamefont {Xiao}},\ and\
  \bibinfo {author} {\bibfnamefont {X.}~\bibnamefont {Xu}},\ }\bibfield
  {title} {\bibinfo {title} {Observation of fractionally quantized anomalous
  hall effect},\ }\href {https://doi.org/10.1038/s41586-023-06536-0} {\bibfield
   {journal} {\bibinfo  {journal} {Nature}\ }\textbf {\bibinfo {volume}
  {622}},\ \bibinfo {pages} {74} (\bibinfo {year} {2023})}\BibitemShut
  {NoStop}%
\bibitem [{\citenamefont {Foutty}\ \emph {et~al.}(2024)\citenamefont {Foutty},
  \citenamefont {Kometter}, \citenamefont {Devakul}, \citenamefont {Reddy},
  \citenamefont {Watanabe}, \citenamefont {Taniguchi}, \citenamefont {Fu},\
  and\ \citenamefont {Feldman}}]{FouttyMappingTwistTunedMultiband2024}%
  \BibitemOpen
  \bibfield  {author} {\bibinfo {author} {\bibfnamefont {B.~A.}\ \bibnamefont
  {Foutty}}, \bibinfo {author} {\bibfnamefont {C.~R.}\ \bibnamefont
  {Kometter}}, \bibinfo {author} {\bibfnamefont {T.}~\bibnamefont {Devakul}},
  \bibinfo {author} {\bibfnamefont {A.~P.}\ \bibnamefont {Reddy}}, \bibinfo
  {author} {\bibfnamefont {K.}~\bibnamefont {Watanabe}}, \bibinfo {author}
  {\bibfnamefont {T.}~\bibnamefont {Taniguchi}}, \bibinfo {author}
  {\bibfnamefont {L.}~\bibnamefont {Fu}},\ and\ \bibinfo {author}
  {\bibfnamefont {B.~E.}\ \bibnamefont {Feldman}},\ }\bibfield  {title}
  {\bibinfo {title} {Mapping twist-tuned multiband topology in bilayer
  {{WSe$_2$}}},\ }\href {https://doi.org/10.1126/science.adi4728} {\bibfield
  {journal} {\bibinfo  {journal} {Science}\ }\textbf {\bibinfo {volume}
  {384}},\ \bibinfo {pages} {343} (\bibinfo {year} {2024})}\BibitemShut
  {NoStop}%
\bibitem [{\citenamefont {Xu}\ \emph {et~al.}(2023)\citenamefont {Xu},
  \citenamefont {Sun}, \citenamefont {Jia}, \citenamefont {Liu}, \citenamefont
  {Xu}, \citenamefont {Li}, \citenamefont {Gu}, \citenamefont {Watanabe},
  \citenamefont {Taniguchi}, \citenamefont {Tong}, \citenamefont {Jia},
  \citenamefont {Shi}, \citenamefont {Jiang}, \citenamefont {Zhang},
  \citenamefont {Liu},\ and\ \citenamefont
  {Li}}]{XuObservationIntegerFractionalQuantumAnomalous2023}%
  \BibitemOpen
  \bibfield  {author} {\bibinfo {author} {\bibfnamefont {F.}~\bibnamefont
  {Xu}}, \bibinfo {author} {\bibfnamefont {Z.}~\bibnamefont {Sun}}, \bibinfo
  {author} {\bibfnamefont {T.}~\bibnamefont {Jia}}, \bibinfo {author}
  {\bibfnamefont {C.}~\bibnamefont {Liu}}, \bibinfo {author} {\bibfnamefont
  {C.}~\bibnamefont {Xu}}, \bibinfo {author} {\bibfnamefont {C.}~\bibnamefont
  {Li}}, \bibinfo {author} {\bibfnamefont {Y.}~\bibnamefont {Gu}}, \bibinfo
  {author} {\bibfnamefont {K.}~\bibnamefont {Watanabe}}, \bibinfo {author}
  {\bibfnamefont {T.}~\bibnamefont {Taniguchi}}, \bibinfo {author}
  {\bibfnamefont {B.}~\bibnamefont {Tong}}, \bibinfo {author} {\bibfnamefont
  {J.}~\bibnamefont {Jia}}, \bibinfo {author} {\bibfnamefont {Z.}~\bibnamefont
  {Shi}}, \bibinfo {author} {\bibfnamefont {S.}~\bibnamefont {Jiang}}, \bibinfo
  {author} {\bibfnamefont {Y.}~\bibnamefont {Zhang}}, \bibinfo {author}
  {\bibfnamefont {X.}~\bibnamefont {Liu}},\ and\ \bibinfo {author}
  {\bibfnamefont {T.}~\bibnamefont {Li}},\ }\bibfield  {title} {\bibinfo
  {title} {Observation of integer and fractional quantum anomalous hall effects
  in twisted bilayer {{MoTe$_2$}}},\ }\href
  {https://doi.org/10.1103/PhysRevX.13.031037} {\bibfield  {journal} {\bibinfo
  {journal} {Phys. Rev. X}\ }\textbf {\bibinfo {volume} {13}},\ \bibinfo
  {pages} {031037} (\bibinfo {year} {2023})}\BibitemShut {NoStop}%
\bibitem [{\citenamefont {Ghiotto}\ \emph {et~al.}(2024)\citenamefont
  {Ghiotto}, \citenamefont {Wei}, \citenamefont {Song}, \citenamefont {Zang},
  \citenamefont {Tazi}, \citenamefont {Ostrom}, \citenamefont {Watanabe},
  \citenamefont {Taniguchi}, \citenamefont {Hone}, \citenamefont {Rhodes},
  \citenamefont {Millis}, \citenamefont {Dean}, \citenamefont {Wang},\ and\
  \citenamefont {Pasupathy}}]{GhiottoStonerInstabilitiesIsing2024}%
  \BibitemOpen
  \bibfield  {author} {\bibinfo {author} {\bibfnamefont {A.}~\bibnamefont
  {Ghiotto}}, \bibinfo {author} {\bibfnamefont {L.}~\bibnamefont {Wei}},
  \bibinfo {author} {\bibfnamefont {L.}~\bibnamefont {Song}}, \bibinfo {author}
  {\bibfnamefont {J.}~\bibnamefont {Zang}}, \bibinfo {author} {\bibfnamefont
  {A.~B.}\ \bibnamefont {Tazi}}, \bibinfo {author} {\bibfnamefont
  {D.}~\bibnamefont {Ostrom}}, \bibinfo {author} {\bibfnamefont
  {K.}~\bibnamefont {Watanabe}}, \bibinfo {author} {\bibfnamefont
  {T.}~\bibnamefont {Taniguchi}}, \bibinfo {author} {\bibfnamefont {J.~C.}\
  \bibnamefont {Hone}}, \bibinfo {author} {\bibfnamefont {D.~A.}\ \bibnamefont
  {Rhodes}}, \bibinfo {author} {\bibfnamefont {A.~J.}\ \bibnamefont {Millis}},
  \bibinfo {author} {\bibfnamefont {C.~R.}\ \bibnamefont {Dean}}, \bibinfo
  {author} {\bibfnamefont {L.}~\bibnamefont {Wang}},\ and\ \bibinfo {author}
  {\bibfnamefont {A.~N.}\ \bibnamefont {Pasupathy}},\ }\href
  {https://arxiv.org/abs/2405.17316} {\bibinfo {title} {Stoner instabilities
  and ising excitonic states in twisted transition metal dichalcogenides}}
  (\bibinfo {year} {2024}),\ \bibinfo {note} {arXiv:2405.17316
  [cond-mat.str-el]}\BibitemShut {NoStop}%
\bibitem [{\citenamefont {Ghiotto}\ \emph {et~al.}(2021)\citenamefont
  {Ghiotto}, \citenamefont {Shih}, \citenamefont {Pereira}, \citenamefont
  {Rhodes}, \citenamefont {Kim}, \citenamefont {Zang}, \citenamefont {Millis},
  \citenamefont {Watanabe}, \citenamefont {Taniguchi}, \citenamefont {Hone},
  \citenamefont {Wang}, \citenamefont {Dean},\ and\ \citenamefont
  {Pasupathy}}]{GhiottoQuantumCriticalityTwisted2021}%
  \BibitemOpen
  \bibfield  {author} {\bibinfo {author} {\bibfnamefont {A.}~\bibnamefont
  {Ghiotto}}, \bibinfo {author} {\bibfnamefont {E.-M.}\ \bibnamefont {Shih}},
  \bibinfo {author} {\bibfnamefont {G.~S. S.~G.}\ \bibnamefont {Pereira}},
  \bibinfo {author} {\bibfnamefont {D.~A.}\ \bibnamefont {Rhodes}}, \bibinfo
  {author} {\bibfnamefont {B.}~\bibnamefont {Kim}}, \bibinfo {author}
  {\bibfnamefont {J.}~\bibnamefont {Zang}}, \bibinfo {author} {\bibfnamefont
  {A.~J.}\ \bibnamefont {Millis}}, \bibinfo {author} {\bibfnamefont
  {K.}~\bibnamefont {Watanabe}}, \bibinfo {author} {\bibfnamefont
  {T.}~\bibnamefont {Taniguchi}}, \bibinfo {author} {\bibfnamefont {J.~C.}\
  \bibnamefont {Hone}}, \bibinfo {author} {\bibfnamefont {L.}~\bibnamefont
  {Wang}}, \bibinfo {author} {\bibfnamefont {C.~R.}\ \bibnamefont {Dean}},\
  and\ \bibinfo {author} {\bibfnamefont {A.~N.}\ \bibnamefont {Pasupathy}},\
  }\bibfield  {title} {\bibinfo {title} {Quantum criticality in twisted
  transition metal dichalcogenides},\ }\href
  {https://doi.org/10.1038/s41586-021-03815-6} {\bibfield  {journal} {\bibinfo
  {journal} {Nature}\ }\textbf {\bibinfo {volume} {597}},\ \bibinfo {pages}
  {345} (\bibinfo {year} {2021})}\BibitemShut {NoStop}%
\bibitem [{\citenamefont {Kiely}\ and\ \citenamefont
  {Chowdhury}(2024)}]{KielyContinuousWigner-MottTransitions2024}%
  \BibitemOpen
  \bibfield  {author} {\bibinfo {author} {\bibfnamefont {T.~G.}\ \bibnamefont
  {Kiely}}\ and\ \bibinfo {author} {\bibfnamefont {D.}~\bibnamefont
  {Chowdhury}},\ }\bibfield  {title} {\bibinfo {title} {Continuous wigner-mott
  transitions at {{$\nu=1/5$}}},\ }\href
  {https://doi.org/10.1103/PhysRevB.110.L241112} {\bibfield  {journal}
  {\bibinfo  {journal} {Phys. Rev. B}\ }\textbf {\bibinfo {volume} {110}},\
  \bibinfo {pages} {L241112} (\bibinfo {year} {2024})}\BibitemShut {NoStop}%
\bibitem [{\citenamefont {Xia}\ \emph {et~al.}(2025)\citenamefont {Xia},
  \citenamefont {Han}, \citenamefont {Watanabe}, \citenamefont {Taniguchi},
  \citenamefont {Shan},\ and\ \citenamefont {Mak}}]{Xia_Nature_833838_2025}%
  \BibitemOpen
  \bibfield  {author} {\bibinfo {author} {\bibfnamefont {Y.}~\bibnamefont
  {Xia}}, \bibinfo {author} {\bibfnamefont {Z.}~\bibnamefont {Han}}, \bibinfo
  {author} {\bibfnamefont {K.}~\bibnamefont {Watanabe}}, \bibinfo {author}
  {\bibfnamefont {T.}~\bibnamefont {Taniguchi}}, \bibinfo {author}
  {\bibfnamefont {J.}~\bibnamefont {Shan}},\ and\ \bibinfo {author}
  {\bibfnamefont {K.~F.}\ \bibnamefont {Mak}},\ }\bibfield  {title} {\bibinfo
  {title} {Superconductivity in twisted bilayer {{WSe$_2$}}},\ }\href
  {https://doi.org/10.1038/s41586-024-08116-2} {\bibfield  {journal} {\bibinfo
  {journal} {Nature}\ }\textbf {\bibinfo {volume} {637}},\ \bibinfo {pages}
  {833} (\bibinfo {year} {2025})}\BibitemShut {NoStop}%
\bibitem [{\citenamefont {Guo}\ \emph {et~al.}(2025)\citenamefont {Guo},
  \citenamefont {Pack}, \citenamefont {Swann}, \citenamefont {Holtzman},
  \citenamefont {Cothrine}, \citenamefont {Watanabe}, \citenamefont
  {Taniguchi}, \citenamefont {Mandrus}, \citenamefont {Barmak}, \citenamefont
  {Hone}, \citenamefont {Millis}, \citenamefont {Pasupathy},\ and\
  \citenamefont {Dean}}]{Guo_Nature_839845_2025}%
  \BibitemOpen
  \bibfield  {author} {\bibinfo {author} {\bibfnamefont {Y.}~\bibnamefont
  {Guo}}, \bibinfo {author} {\bibfnamefont {J.}~\bibnamefont {Pack}}, \bibinfo
  {author} {\bibfnamefont {J.}~\bibnamefont {Swann}}, \bibinfo {author}
  {\bibfnamefont {L.}~\bibnamefont {Holtzman}}, \bibinfo {author}
  {\bibfnamefont {M.}~\bibnamefont {Cothrine}}, \bibinfo {author}
  {\bibfnamefont {K.}~\bibnamefont {Watanabe}}, \bibinfo {author}
  {\bibfnamefont {T.}~\bibnamefont {Taniguchi}}, \bibinfo {author}
  {\bibfnamefont {D.~G.}\ \bibnamefont {Mandrus}}, \bibinfo {author}
  {\bibfnamefont {K.}~\bibnamefont {Barmak}}, \bibinfo {author} {\bibfnamefont
  {J.}~\bibnamefont {Hone}}, \bibinfo {author} {\bibfnamefont {A.~J.}\
  \bibnamefont {Millis}}, \bibinfo {author} {\bibfnamefont {A.}~\bibnamefont
  {Pasupathy}},\ and\ \bibinfo {author} {\bibfnamefont {C.~R.}\ \bibnamefont
  {Dean}},\ }\bibfield  {title} {\bibinfo {title} {Superconductivity in 5.0°
  twisted bilayer {{WSe$_2$}}},\ }\href
  {https://doi.org/10.1038/s41586-024-08381-1} {\bibfield  {journal} {\bibinfo
  {journal} {Nature}\ }\textbf {\bibinfo {volume} {637}},\ \bibinfo {pages}
  {839} (\bibinfo {year} {2025})}\BibitemShut {NoStop}%
\bibitem [{\citenamefont {{Kn{\"u}ppel, Patrick and Zhu, Jiacheng and Xia, Yiyu
  and Xia, Zhengchao and Han, Zhongdong and Zeng, Yihang and Watanabe, Kenji
  and Taniguchi, Takashi and Shan, Jie and Mak, Kin
  Fai}}(2025)}]{Knuppel_NatComm_2025}%
  \BibitemOpen
  \bibfield  {author} {\bibinfo {author} {\bibnamefont {{Kn{\"u}ppel, Patrick
  and Zhu, Jiacheng and Xia, Yiyu and Xia, Zhengchao and Han, Zhongdong and
  Zeng, Yihang and Watanabe, Kenji and Taniguchi, Takashi and Shan, Jie and
  Mak, Kin Fai}}},\ }\bibfield  {title} {\bibinfo {title} {Correlated states
  controlled by a tunable van hove singularity in moir{\'e} wse$_2$ bilayers},\
  }\href {https://doi.org/10.1038/s41467-025-57235-5} {\bibfield  {journal}
  {\bibinfo  {journal} {Nat. Commun.}\ }\textbf {\bibinfo {volume} {16}},\
  \bibinfo {pages} {1959} (\bibinfo {year} {2025})}\BibitemShut {NoStop}%
\bibitem [{\citenamefont {Xia}\ \emph {et~al.}(2026)\citenamefont {Xia},
  \citenamefont {Han}, \citenamefont {Zhu}, \citenamefont {Zhang},
  \citenamefont {Kn{\"u}ppel}, \citenamefont {Watanabe}, \citenamefont
  {Taniguchi}, \citenamefont {Mak},\ and\ \citenamefont
  {Shan}}]{Xia_BandwidthTunedMottTransitionSuperconductivityMoireWSe2_2026}%
  \BibitemOpen
  \bibfield  {author} {\bibinfo {author} {\bibfnamefont {Y.}~\bibnamefont
  {Xia}}, \bibinfo {author} {\bibfnamefont {Z.}~\bibnamefont {Han}}, \bibinfo
  {author} {\bibfnamefont {J.}~\bibnamefont {Zhu}}, \bibinfo {author}
  {\bibfnamefont {Y.}~\bibnamefont {Zhang}}, \bibinfo {author} {\bibfnamefont
  {P.}~\bibnamefont {Kn{\"u}ppel}}, \bibinfo {author} {\bibfnamefont
  {K.}~\bibnamefont {Watanabe}}, \bibinfo {author} {\bibfnamefont
  {T.}~\bibnamefont {Taniguchi}}, \bibinfo {author} {\bibfnamefont {K.~F.}\
  \bibnamefont {Mak}},\ and\ \bibinfo {author} {\bibfnamefont {J.}~\bibnamefont
  {Shan}},\ }\bibfield  {title} {\bibinfo {title} {Bandwidth-tuned mott
  transition and superconductivity in moir{\'e} wse$_2$},\ }\href
  {https://doi.org/10.1038/s41586-025-10049-3} {\bibfield  {journal} {\bibinfo
  {journal} {Nature}\ }\textbf {\bibinfo {volume} {650}},\ \bibinfo {pages}
  {585} (\bibinfo {year} {2026})}\BibitemShut {NoStop}%
\bibitem [{\citenamefont {Guo}\ \emph {et~al.}(2026)\citenamefont {Guo},
  \citenamefont {Cenker}, \citenamefont {Fischer}, \citenamefont
  {Muñoz-Segovia}, \citenamefont {Pack}, \citenamefont {Holtzman},
  \citenamefont {Klebl}, \citenamefont {Watanabe}, \citenamefont {Taniguchi},
  \citenamefont {Barmak}, \citenamefont {Hone}, \citenamefont {Rubio},
  \citenamefont {Kennes}, \citenamefont {Millis}, \citenamefont {Pasupathy},\
  and\ \citenamefont
  {Dean}}]{Guo_AngleEvolutionSuperconductingPhaseDiagram_tWSe2_Nature_2026}%
  \BibitemOpen
  \bibfield  {author} {\bibinfo {author} {\bibfnamefont {Y.}~\bibnamefont
  {Guo}}, \bibinfo {author} {\bibfnamefont {J.}~\bibnamefont {Cenker}},
  \bibinfo {author} {\bibfnamefont {A.}~\bibnamefont {Fischer}}, \bibinfo
  {author} {\bibfnamefont {D.}~\bibnamefont {Muñoz-Segovia}}, \bibinfo
  {author} {\bibfnamefont {J.}~\bibnamefont {Pack}}, \bibinfo {author}
  {\bibfnamefont {L.}~\bibnamefont {Holtzman}}, \bibinfo {author}
  {\bibfnamefont {L.}~\bibnamefont {Klebl}}, \bibinfo {author} {\bibfnamefont
  {K.}~\bibnamefont {Watanabe}}, \bibinfo {author} {\bibfnamefont
  {T.}~\bibnamefont {Taniguchi}}, \bibinfo {author} {\bibfnamefont
  {K.}~\bibnamefont {Barmak}}, \bibinfo {author} {\bibfnamefont
  {J.}~\bibnamefont {Hone}}, \bibinfo {author} {\bibfnamefont {A.}~\bibnamefont
  {Rubio}}, \bibinfo {author} {\bibfnamefont {D.~M.}\ \bibnamefont {Kennes}},
  \bibinfo {author} {\bibfnamefont {A.~J.}\ \bibnamefont {Millis}}, \bibinfo
  {author} {\bibfnamefont {A.}~\bibnamefont {Pasupathy}},\ and\ \bibinfo
  {author} {\bibfnamefont {C.~R.}\ \bibnamefont {Dean}},\ }\bibfield  {title}
  {\bibinfo {title} {Angle evolution of the superconducting phase diagram in
  twisted bilayer wse2},\ }\href {https://doi.org/10.1038/s41586-026-10357-2}
  {\bibfield  {journal} {\bibinfo  {journal} {Nature}\ }\textbf {\bibinfo
  {volume} {652}},\ \bibinfo {pages} {622} (\bibinfo {year}
  {2026})}\BibitemShut {NoStop}%
\bibitem [{\citenamefont {Pan}\ \emph {et~al.}(2020)\citenamefont {Pan},
  \citenamefont {Wu},\ and\ \citenamefont
  {Das~Sarma}}]{PanBandTopologyHubbard2020}%
  \BibitemOpen
  \bibfield  {author} {\bibinfo {author} {\bibfnamefont {H.}~\bibnamefont
  {Pan}}, \bibinfo {author} {\bibfnamefont {F.}~\bibnamefont {Wu}},\ and\
  \bibinfo {author} {\bibfnamefont {S.}~\bibnamefont {Das~Sarma}},\ }\bibfield
  {title} {\bibinfo {title} {Band topology, hubbard model, heisenberg model,
  and dzyaloshinskii-moriya interaction in twisted bilayer {{WSe$_2$}}},\
  }\href {https://doi.org/10.1103/PhysRevResearch.2.033087} {\bibfield
  {journal} {\bibinfo  {journal} {Phys. Rev. Res.}\ }\textbf {\bibinfo {volume}
  {2}},\ \bibinfo {pages} {033087} (\bibinfo {year} {2020})}\BibitemShut
  {NoStop}%
\bibitem [{\citenamefont {Kormányos}\ \emph {et~al.}(2015)\citenamefont
  {Kormányos}, \citenamefont {Burkard}, \citenamefont {Gmitra}, \citenamefont
  {Fabian}, \citenamefont {Zólyomi}, \citenamefont {Drummond},\ and\
  \citenamefont {Fal’ko}}]{Kormányos_kpTheoryTwoDimensional2015}%
  \BibitemOpen
  \bibfield  {author} {\bibinfo {author} {\bibfnamefont {A.}~\bibnamefont
  {Kormányos}}, \bibinfo {author} {\bibfnamefont {G.}~\bibnamefont {Burkard}},
  \bibinfo {author} {\bibfnamefont {M.}~\bibnamefont {Gmitra}}, \bibinfo
  {author} {\bibfnamefont {J.}~\bibnamefont {Fabian}}, \bibinfo {author}
  {\bibfnamefont {V.}~\bibnamefont {Zólyomi}}, \bibinfo {author}
  {\bibfnamefont {N.~D.}\ \bibnamefont {Drummond}},\ and\ \bibinfo {author}
  {\bibfnamefont {V.}~\bibnamefont {Fal’ko}},\ }\bibfield  {title} {\bibinfo
  {title} {{{$\boldsymbol{k}\cdot\boldsymbol{p}$}} theory for two-dimensional
  transition metal dichalcogenide semiconductors},\ }\href
  {https://doi.org/10.1088/2053-1583/2/2/022001} {\bibfield  {journal}
  {\bibinfo  {journal} {2D Materials}\ }\textbf {\bibinfo {volume} {2}},\
  \bibinfo {pages} {022001} (\bibinfo {year} {2015})}\BibitemShut {NoStop}%
\bibitem [{\citenamefont {Devakul}\ \emph {et~al.}(2021)\citenamefont
  {Devakul}, \citenamefont {Crépel}, \citenamefont {Zhang},\ and\
  \citenamefont {Fu}}]{DevakulMagicInTwistedTransition2021}%
  \BibitemOpen
  \bibfield  {author} {\bibinfo {author} {\bibfnamefont {T.}~\bibnamefont
  {Devakul}}, \bibinfo {author} {\bibfnamefont {V.}~\bibnamefont {Crépel}},
  \bibinfo {author} {\bibfnamefont {Y.}~\bibnamefont {Zhang}},\ and\ \bibinfo
  {author} {\bibfnamefont {L.}~\bibnamefont {Fu}},\ }\bibfield  {title}
  {\bibinfo {title} {Magic in twisted transition metal dichalcogenide
  bilayers},\ }\href {https://doi.org/10.1038/s41467-021-27042-9} {\bibfield
  {journal} {\bibinfo  {journal} {Nat. Comm.}\ }\textbf {\bibinfo {volume}
  {12}},\ \bibinfo {pages} {6730} (\bibinfo {year} {2021})}\BibitemShut
  {NoStop}%
\bibitem [{\citenamefont {Crépel}\ and\ \citenamefont
  {Millis}(2024{\natexlab{a}})}]{CrepelBridgingSmallLargeTwisted2024}%
  \BibitemOpen
  \bibfield  {author} {\bibinfo {author} {\bibfnamefont {V.}~\bibnamefont
  {Crépel}}\ and\ \bibinfo {author} {\bibfnamefont {A.}~\bibnamefont
  {Millis}},\ }\bibfield  {title} {\bibinfo {title} {Bridging the small and
  large in twisted transition metal dichalcogenide homobilayers: A
  tight-binding model capturing orbital interference and topology across a wide
  range of twist angles},\ }\href
  {https://doi.org/10.1103/PhysRevResearch.6.033127} {\bibfield  {journal}
  {\bibinfo  {journal} {Phys. Rev. Res.}\ }\textbf {\bibinfo {volume} {6}},\
  \bibinfo {pages} {033127} (\bibinfo {year} {2024}{\natexlab{a}})}\BibitemShut
  {NoStop}%
\bibitem [{\citenamefont {Zhang}\ \emph {et~al.}(2024)\citenamefont {Zhang},
  \citenamefont {Pi}, \citenamefont {Liu}, \citenamefont {Miao}, \citenamefont
  {Qi}, \citenamefont {Regnault}, \citenamefont {Weng}, \citenamefont {Dai},
  \citenamefont {Bernevig}, \citenamefont {Wu},\ and\ \citenamefont
  {Yu}}]{ZhangUniversalMoire-Model-Building2024}%
  \BibitemOpen
  \bibfield  {author} {\bibinfo {author} {\bibfnamefont {Y.}~\bibnamefont
  {Zhang}}, \bibinfo {author} {\bibfnamefont {H.}~\bibnamefont {Pi}}, \bibinfo
  {author} {\bibfnamefont {J.}~\bibnamefont {Liu}}, \bibinfo {author}
  {\bibfnamefont {W.}~\bibnamefont {Miao}}, \bibinfo {author} {\bibfnamefont
  {Z.}~\bibnamefont {Qi}}, \bibinfo {author} {\bibfnamefont {N.}~\bibnamefont
  {Regnault}}, \bibinfo {author} {\bibfnamefont {H.}~\bibnamefont {Weng}},
  \bibinfo {author} {\bibfnamefont {X.}~\bibnamefont {Dai}}, \bibinfo {author}
  {\bibfnamefont {B.~A.}\ \bibnamefont {Bernevig}}, \bibinfo {author}
  {\bibfnamefont {Q.}~\bibnamefont {Wu}},\ and\ \bibinfo {author}
  {\bibfnamefont {J.}~\bibnamefont {Yu}},\ }\href
  {https://arxiv.org/abs/2411.08108} {\bibinfo {title} {Universal
  moiré-model-building method without fitting: Application to twisted
  {{MoTe$_2$}} and {{WSe$_2$}}}} (\bibinfo {year} {2024}),\ \bibinfo {note}
  {arXiv:2411.08108 [cond-mat.mes-hall]}\BibitemShut {NoStop}%
\bibitem [{\citenamefont {Akbar}\ \emph {et~al.}(2024)\citenamefont {Akbar},
  \citenamefont {Biborski}, \citenamefont {Rademaker},\ and\ \citenamefont
  {Zegrodnik}}]{AkbarTopologicalSuperconductivityMixed2024}%
  \BibitemOpen
  \bibfield  {author} {\bibinfo {author} {\bibfnamefont {W.}~\bibnamefont
  {Akbar}}, \bibinfo {author} {\bibfnamefont {A.}~\bibnamefont {Biborski}},
  \bibinfo {author} {\bibfnamefont {L.}~\bibnamefont {Rademaker}},\ and\
  \bibinfo {author} {\bibfnamefont {M.}~\bibnamefont {Zegrodnik}},\ }\bibfield
  {title} {\bibinfo {title} {Topological superconductivity with mixed
  singlet-triplet pairing in moiré transition metal dichalcogenide bilayers},\
  }\href {https://doi.org/10.1103/PhysRevB.110.064516} {\bibfield  {journal}
  {\bibinfo  {journal} {Phys. Rev. B}\ }\textbf {\bibinfo {volume} {110}},\
  \bibinfo {pages} {064516} (\bibinfo {year} {2024})}\BibitemShut {NoStop}%
\bibitem [{\citenamefont {Zegrodnik}\ and\ \citenamefont
  {Biborski}(2023)}]{ZegrodnikMixedSinglet-TripletSuperconductingState2023}%
  \BibitemOpen
  \bibfield  {author} {\bibinfo {author} {\bibfnamefont {M.}~\bibnamefont
  {Zegrodnik}}\ and\ \bibinfo {author} {\bibfnamefont {A.}~\bibnamefont
  {Biborski}},\ }\bibfield  {title} {\bibinfo {title} {Mixed singlet-triplet
  superconducting state within the moiré model applied to twisted bilayer},\
  }\href {https://doi.org/10.1103/PhysRevB.108.064506} {\bibfield  {journal}
  {\bibinfo  {journal} {Phys. Rev. B}\ }\textbf {\bibinfo {volume} {108}},\
  \bibinfo {pages} {064506} (\bibinfo {year} {2023})}\BibitemShut {NoStop}%
\bibitem [{\citenamefont {Wu}\ \emph {et~al.}(2023)\citenamefont {Wu},
  \citenamefont {Wu},\ and\ \citenamefont
  {Yao}}]{WuPair-Density-WaveChiralSuperconductivity2023}%
  \BibitemOpen
  \bibfield  {author} {\bibinfo {author} {\bibfnamefont {Y.-M.}\ \bibnamefont
  {Wu}}, \bibinfo {author} {\bibfnamefont {Z.}~\bibnamefont {Wu}},\ and\
  \bibinfo {author} {\bibfnamefont {H.}~\bibnamefont {Yao}},\ }\bibfield
  {title} {\bibinfo {title} {Pair-density-wave and chiral superconductivity in
  twisted bilayer transition metal dichalcogenides},\ }\href
  {https://doi.org/10.1103/PhysRevLett.130.126001} {\bibfield  {journal}
  {\bibinfo  {journal} {Phys. Rev. Lett.}\ }\textbf {\bibinfo {volume} {130}},\
  \bibinfo {pages} {126001} (\bibinfo {year} {2023})}\BibitemShut {NoStop}%
\bibitem [{\citenamefont {Kim}\ \emph {et~al.}(2025)\citenamefont {Kim},
  \citenamefont {Mendez‐Valderrama}, \citenamefont {Wang},\ and\
  \citenamefont {Chowdhury}}]{KimTheoryCorrelatedInsulatorsSuperconductor2025}%
  \BibitemOpen
  \bibfield  {author} {\bibinfo {author} {\bibfnamefont {S.}~\bibnamefont
  {Kim}}, \bibinfo {author} {\bibfnamefont {J.~F.}\ \bibnamefont
  {Mendez‐Valderrama}}, \bibinfo {author} {\bibfnamefont {X.}~\bibnamefont
  {Wang}},\ and\ \bibinfo {author} {\bibfnamefont {D.}~\bibnamefont
  {Chowdhury}},\ }\bibfield  {title} {\bibinfo {title} {Theory of correlated
  insulators and superconductor at {{$\nu=1$}} in twisted {{WSe$_2$}}},\ }\href
  {https://doi.org/10.1038/s41467-025-56816-8} {\bibfield  {journal} {\bibinfo
  {journal} {Nat. Comm.}\ }\textbf {\bibinfo {volume} {16}},\ \bibinfo {pages}
  {1701} (\bibinfo {year} {2025})}\BibitemShut {NoStop}%
\bibitem [{\citenamefont {Abouelkomsan}\ \emph {et~al.}(2024)\citenamefont
  {Abouelkomsan}, \citenamefont {Bergholtz},\ and\ \citenamefont
  {Chatterjee}}]{AbouelkomsanMultiferroicityTopologyTwisted2024}%
  \BibitemOpen
  \bibfield  {author} {\bibinfo {author} {\bibfnamefont {A.}~\bibnamefont
  {Abouelkomsan}}, \bibinfo {author} {\bibfnamefont {E.~J.}\ \bibnamefont
  {Bergholtz}},\ and\ \bibinfo {author} {\bibfnamefont {S.}~\bibnamefont
  {Chatterjee}},\ }\bibfield  {title} {\bibinfo {title} {Multiferroicity and
  topology in twisted transition metal dichalcogenides},\ }\href
  {https://doi.org/10.1103/PhysRevLett.133.026801} {\bibfield  {journal}
  {\bibinfo  {journal} {Phys. Rev. Lett.}\ }\textbf {\bibinfo {volume} {133}},\
  \bibinfo {pages} {026801} (\bibinfo {year} {2024})}\BibitemShut {NoStop}%
\bibitem [{\citenamefont {Hsu}\ \emph {et~al.}(2021)\citenamefont {Hsu},
  \citenamefont {Wu},\ and\ \citenamefont
  {Das~Sarma}}]{HsuSpin-valleyLockedInstabilities2021}%
  \BibitemOpen
  \bibfield  {author} {\bibinfo {author} {\bibfnamefont {Y.-T.}\ \bibnamefont
  {Hsu}}, \bibinfo {author} {\bibfnamefont {F.}~\bibnamefont {Wu}},\ and\
  \bibinfo {author} {\bibfnamefont {S.}~\bibnamefont {Das~Sarma}},\ }\bibfield
  {title} {\bibinfo {title} {Spin-valley locked instabilities in moiré
  transition metal dichalcogenides with conventional and higher-order van hove
  singularities},\ }\href {https://doi.org/10.1103/PhysRevB.104.195134}
  {\bibfield  {journal} {\bibinfo  {journal} {Phys. Rev. B}\ }\textbf {\bibinfo
  {volume} {104}},\ \bibinfo {pages} {195134} (\bibinfo {year}
  {2021})}\BibitemShut {NoStop}%
\bibitem [{\citenamefont {Zhu}\ \emph {et~al.}(2025)\citenamefont {Zhu},
  \citenamefont {Chou}, \citenamefont {Xie},\ and\ \citenamefont
  {Das~Sarma}}]{ZhuSuperconductivityInTwistedTransition2025}%
  \BibitemOpen
  \bibfield  {author} {\bibinfo {author} {\bibfnamefont {J.}~\bibnamefont
  {Zhu}}, \bibinfo {author} {\bibfnamefont {Y.-Z.}\ \bibnamefont {Chou}},
  \bibinfo {author} {\bibfnamefont {M.}~\bibnamefont {Xie}},\ and\ \bibinfo
  {author} {\bibfnamefont {S.}~\bibnamefont {Das~Sarma}},\ }\bibfield  {title}
  {\bibinfo {title} {Superconductivity in twisted transition metal
  dichalcogenide homobilayers},\ }\href
  {https://doi.org/10.1103/PhysRevB.111.L060501} {\bibfield  {journal}
  {\bibinfo  {journal} {Phys. Rev. B}\ }\textbf {\bibinfo {volume} {111}},\
  \bibinfo {pages} {L060501} (\bibinfo {year} {2025})}\BibitemShut {NoStop}%
\bibitem [{\citenamefont {Crépel}\ and\ \citenamefont
  {Millis}(2024{\natexlab{b}})}]{CrépelMillisSpinonPairingInduced2024}%
  \BibitemOpen
  \bibfield  {author} {\bibinfo {author} {\bibfnamefont {V.}~\bibnamefont
  {Crépel}}\ and\ \bibinfo {author} {\bibfnamefont {A.}~\bibnamefont
  {Millis}},\ }\bibfield  {title} {\bibinfo {title} {Spinon pairing induced by
  chiral in-plane exchange and the stabilization of odd-spin chern number spin
  liquid in twisted mote$_2$},\ }\href
  {https://doi.org/10.1103/PhysRevLett.133.146503} {\bibfield  {journal}
  {\bibinfo  {journal} {Phys. Rev. Lett.}\ }\textbf {\bibinfo {volume} {133}},\
  \bibinfo {pages} {146503} (\bibinfo {year} {2024}{\natexlab{b}})}\BibitemShut
  {NoStop}%
\bibitem [{\citenamefont {Bélanger}\ \emph {et~al.}(2022)\citenamefont
  {Bélanger}, \citenamefont {Fournier},\ and\ \citenamefont
  {Sénéchal}}]{BelangerSuperconductivityTwistedBilayerTransitionMetalDichalcogenideQuantumCluster2022}%
  \BibitemOpen
  \bibfield  {author} {\bibinfo {author} {\bibfnamefont {M.}~\bibnamefont
  {Bélanger}}, \bibinfo {author} {\bibfnamefont {J.}~\bibnamefont
  {Fournier}},\ and\ \bibinfo {author} {\bibfnamefont {D.}~\bibnamefont
  {Sénéchal}},\ }\bibfield  {title} {\bibinfo {title} {Superconductivity in
  the twisted bilayer transition-metal dichalcogenide: A quantum cluster
  study},\ }\href {https://doi.org/10.1103/PhysRevB.106.235135} {\bibfield
  {journal} {\bibinfo  {journal} {Phys. Rev. B}\ }\textbf {\bibinfo {volume}
  {106}},\ \bibinfo {pages} {235135} (\bibinfo {year} {2022})}\BibitemShut
  {NoStop}%
\bibitem [{\citenamefont {Chen}\ and\ \citenamefont
  {Sheng}(2023)}]{ChenSingletTripletPairDensityWave2023}%
  \BibitemOpen
  \bibfield  {author} {\bibinfo {author} {\bibfnamefont {F.}~\bibnamefont
  {Chen}}\ and\ \bibinfo {author} {\bibfnamefont {D.~N.}\ \bibnamefont
  {Sheng}},\ }\bibfield  {title} {\bibinfo {title} {Singlet, triplet, and pair
  density wave superconductivity in the doped triangular-lattice moiré
  system},\ }\href {https://doi.org/10.1103/PhysRevB.108.L201110} {\bibfield
  {journal} {\bibinfo  {journal} {Phys. Rev. B}\ }\textbf {\bibinfo {volume}
  {108}},\ \bibinfo {pages} {L201110} (\bibinfo {year} {2023})}\BibitemShut
  {NoStop}%
\bibitem [{\citenamefont {Zhou}\ and\ \citenamefont
  {Zhang}(2023)}]{ZhouModelValleyContrastingFlux2023}%
  \BibitemOpen
  \bibfield  {author} {\bibinfo {author} {\bibfnamefont {B.}~\bibnamefont
  {Zhou}}\ and\ \bibinfo {author} {\bibfnamefont {Y.-H.}\ \bibnamefont
  {Zhang}},\ }\bibfield  {title} {\bibinfo {title} {Chiral and nodal
  superconductors in the {{$t-J$}} model with valley contrasting flux on a
  triangular moiré lattice},\ }\href
  {https://doi.org/10.1103/PhysRevB.108.155111} {\bibfield  {journal} {\bibinfo
   {journal} {Phys. Rev. B}\ }\textbf {\bibinfo {volume} {108}},\ \bibinfo
  {pages} {155111} (\bibinfo {year} {2023})}\BibitemShut {NoStop}%
\bibitem [{\citenamefont {Fischer}\ \emph {et~al.}(2025)\citenamefont
  {Fischer}, \citenamefont {Klebl}, \citenamefont {Cr\'epel}, \citenamefont
  {Ryee}, \citenamefont {Rubio}, \citenamefont {Xian}, \citenamefont {Wehling},
  \citenamefont {Georges}, \citenamefont {Kennes},\ and\ \citenamefont
  {Millis}}]{Fischer_Phys.Rev.X_041055_2025}%
  \BibitemOpen
  \bibfield  {author} {\bibinfo {author} {\bibfnamefont {A.}~\bibnamefont
  {Fischer}}, \bibinfo {author} {\bibfnamefont {L.}~\bibnamefont {Klebl}},
  \bibinfo {author} {\bibfnamefont {V.}~\bibnamefont {Cr\'epel}}, \bibinfo
  {author} {\bibfnamefont {S.}~\bibnamefont {Ryee}}, \bibinfo {author}
  {\bibfnamefont {A.}~\bibnamefont {Rubio}}, \bibinfo {author} {\bibfnamefont
  {L.}~\bibnamefont {Xian}}, \bibinfo {author} {\bibfnamefont {T.~O.}\
  \bibnamefont {Wehling}}, \bibinfo {author} {\bibfnamefont {A.}~\bibnamefont
  {Georges}}, \bibinfo {author} {\bibfnamefont {D.~M.}\ \bibnamefont
  {Kennes}},\ and\ \bibinfo {author} {\bibfnamefont {A.~J.}\ \bibnamefont
  {Millis}},\ }\bibfield  {title} {\bibinfo {title} {Theory of
  intervalley-coherent afm order and topological superconductivity in
  ${\mathrm{twse}}_{2}$},\ }\href {https://doi.org/10.1103/gfzx-rrcr}
  {\bibfield  {journal} {\bibinfo  {journal} {Phys. Rev. X}\ }\textbf {\bibinfo
  {volume} {15}},\ \bibinfo {pages} {041055} (\bibinfo {year}
  {2025})}\BibitemShut {NoStop}%
\bibitem [{\citenamefont {Tuo}\ \emph {et~al.}(2025)\citenamefont {Tuo},
  \citenamefont {Li}, \citenamefont {Wu}, \citenamefont {Sun},\ and\
  \citenamefont
  {Yao}}]{TuoTheoryTopologicalSuperconductivityAntiferromagnetic2025}%
  \BibitemOpen
  \bibfield  {author} {\bibinfo {author} {\bibfnamefont {C.}~\bibnamefont
  {Tuo}}, \bibinfo {author} {\bibfnamefont {M.-R.}\ \bibnamefont {Li}},
  \bibinfo {author} {\bibfnamefont {Z.}~\bibnamefont {Wu}}, \bibinfo {author}
  {\bibfnamefont {W.}~\bibnamefont {Sun}},\ and\ \bibinfo {author}
  {\bibfnamefont {H.}~\bibnamefont {Yao}},\ }\bibfield  {title} {\bibinfo
  {title} {Theory of topological superconductivity and antiferromagnetic
  correlated insulators in twisted bilayer {{WSe$_2$}}},\ }\href
  {https://doi.org/10.1038/s41467-025-64519-3} {\bibfield  {journal} {\bibinfo
  {journal} {Nat. Comm.}\ }\textbf {\bibinfo {volume} {16}},\ \bibinfo {pages}
  {9525} (\bibinfo {year} {2025})}\BibitemShut {NoStop}%
\bibitem [{\citenamefont {Guerci}\ \emph {et~al.}(2024)\citenamefont {Guerci},
  \citenamefont {Kaplan}, \citenamefont {Ingham}, \citenamefont {Pixley},\ and\
  \citenamefont {Millis}}]{GuerciTopologicalSuperconductivityRepulsive2024}%
  \BibitemOpen
  \bibfield  {author} {\bibinfo {author} {\bibfnamefont {D.}~\bibnamefont
  {Guerci}}, \bibinfo {author} {\bibfnamefont {D.}~\bibnamefont {Kaplan}},
  \bibinfo {author} {\bibfnamefont {J.}~\bibnamefont {Ingham}}, \bibinfo
  {author} {\bibfnamefont {J.~H.}\ \bibnamefont {Pixley}},\ and\ \bibinfo
  {author} {\bibfnamefont {A.~J.}\ \bibnamefont {Millis}},\ }\href
  {https://arxiv.org/abs/2408.16075} {\bibinfo {title} {Topological
  superconductivity from repulsive interactions in twisted wse$_2$}} (\bibinfo
  {year} {2024}),\ \bibinfo {note} {arXiv:2408.16075
  [cond-mat.supr-con]}\BibitemShut {NoStop}%
\bibitem [{\citenamefont {Xie}\ and\ \citenamefont
  {Law}(2023)}]{XieOrbitalFuldeFerrellPairing2023}%
  \BibitemOpen
  \bibfield  {author} {\bibinfo {author} {\bibfnamefont {Y.-M.}\ \bibnamefont
  {Xie}}\ and\ \bibinfo {author} {\bibfnamefont {K.~T.}\ \bibnamefont {Law}},\
  }\bibfield  {title} {\bibinfo {title} {Orbital fulde-ferrell pairing state in
  moiré ising superconductors},\ }\href
  {https://doi.org/10.1103/PhysRevLett.131.016001} {\bibfield  {journal}
  {\bibinfo  {journal} {Phys. Rev. Lett.}\ }\textbf {\bibinfo {volume} {131}},\
  \bibinfo {pages} {016001} (\bibinfo {year} {2023})}\BibitemShut {NoStop}%
\bibitem [{\citenamefont {Mu\~noz Segovia}\ \emph {et~al.}(2025)\citenamefont
  {Mu\~noz Segovia}, \citenamefont {Cr\'epel}, \citenamefont {Queiroz},\ and\
  \citenamefont {Millis}}]{MunozSegovia_Phys.Rev.B_085111_2025}%
  \BibitemOpen
  \bibfield  {author} {\bibinfo {author} {\bibfnamefont {D.}~\bibnamefont
  {Mu\~noz Segovia}}, \bibinfo {author} {\bibfnamefont {V.}~\bibnamefont
  {Cr\'epel}}, \bibinfo {author} {\bibfnamefont {R.}~\bibnamefont {Queiroz}},\
  and\ \bibinfo {author} {\bibfnamefont {A.~J.}\ \bibnamefont {Millis}},\
  }\bibfield  {title} {\bibinfo {title} {Twist-angle evolution of the
  intervalley-coherent antiferromagnet in twisted ${\mathrm{wse}}_{2}$},\
  }\href {https://doi.org/10.1103/m6c2-yd5j} {\bibfield  {journal} {\bibinfo
  {journal} {Phys. Rev. B}\ }\textbf {\bibinfo {volume} {112}},\ \bibinfo
  {pages} {085111} (\bibinfo {year} {2025})}\BibitemShut {NoStop}%
\bibitem [{\citenamefont {Schrade}\ and\ \citenamefont
  {Fu}(2024)}]{SchradeFuNematicChiralTopologicalSuperconductivity2024}%
  \BibitemOpen
  \bibfield  {author} {\bibinfo {author} {\bibfnamefont {C.}~\bibnamefont
  {Schrade}}\ and\ \bibinfo {author} {\bibfnamefont {L.}~\bibnamefont {Fu}},\
  }\bibfield  {title} {\bibinfo {title} {Nematic, chiral, and topological
  superconductivity in twisted transition-metal dichalcogenides},\ }\href
  {https://doi.org/10.1103/PhysRevB.110.035143} {\bibfield  {journal} {\bibinfo
   {journal} {Phys. Rev. B}\ }\textbf {\bibinfo {volume} {110}},\ \bibinfo
  {pages} {035143} (\bibinfo {year} {2024})}\BibitemShut {NoStop}%
\bibitem [{\citenamefont {Graser}\ \emph {et~al.}(2009)\citenamefont {Graser},
  \citenamefont {Maier}, \citenamefont {Hirschfeld},\ and\ \citenamefont
  {ino}}]{graserNeardegeneracySeveralPairing2009}%
  \BibitemOpen
  \bibfield  {author} {\bibinfo {author} {\bibfnamefont {S.}~\bibnamefont
  {Graser}}, \bibinfo {author} {\bibfnamefont {T.~A.}\ \bibnamefont {Maier}},
  \bibinfo {author} {\bibfnamefont {P.~J.}\ \bibnamefont {Hirschfeld}},\ and\
  \bibinfo {author} {\bibfnamefont {D.~J.}\ \bibnamefont {ino}},\ }\bibfield
  {title} {\bibinfo {title} {Near-degeneracy of several pairing channels in
  multiorbital models for the {{Fe}} pnictides},\ }\href
  {https://doi.org/10.1088/1367-2630/11/2/025016} {\bibfield  {journal}
  {\bibinfo  {journal} {New J. Phys.}\ }\textbf {\bibinfo {volume} {11}},\
  \bibinfo {pages} {025016} (\bibinfo {year} {2009})}\BibitemShut {NoStop}%
\bibitem [{\citenamefont {Altmeyer}\ \emph {et~al.}(2016)\citenamefont
  {Altmeyer}, \citenamefont {Guterding}, \citenamefont {Hirschfeld},
  \citenamefont {Maier}, \citenamefont {Valent{\'i}},\ and\ \citenamefont
  {Scalapino}}]{altmeyerRoleVertexCorrections2016}%
  \BibitemOpen
  \bibfield  {author} {\bibinfo {author} {\bibfnamefont {M.}~\bibnamefont
  {Altmeyer}}, \bibinfo {author} {\bibfnamefont {D.}~\bibnamefont {Guterding}},
  \bibinfo {author} {\bibfnamefont {P.~J.}\ \bibnamefont {Hirschfeld}},
  \bibinfo {author} {\bibfnamefont {T.~A.}\ \bibnamefont {Maier}}, \bibinfo
  {author} {\bibfnamefont {R.}~\bibnamefont {Valent{\'i}}},\ and\ \bibinfo
  {author} {\bibfnamefont {D.~J.}\ \bibnamefont {Scalapino}},\ }\bibfield
  {title} {\bibinfo {title} {Role of vertex corrections in the matrix
  formulation of the random phase approximation for the multiorbital
  {{Hubbard}} model},\ }\href {https://doi.org/10.1103/PhysRevB.94.214515}
  {\bibfield  {journal} {\bibinfo  {journal} {Phys. Rev. B}\ }\textbf {\bibinfo
  {volume} {94}},\ \bibinfo {pages} {214515} (\bibinfo {year}
  {2016})}\BibitemShut {NoStop}%
\bibitem [{\citenamefont {Esirgen}\ \emph {et~al.}(1999)\citenamefont
  {Esirgen}, \citenamefont {Sch{\"u}ttler},\ and\ \citenamefont
  {Bickers}}]{esirgenMathitWavePairing1999a}%
  \BibitemOpen
  \bibfield  {author} {\bibinfo {author} {\bibfnamefont {G.}~\bibnamefont
  {Esirgen}}, \bibinfo {author} {\bibfnamefont {H.-B.}\ \bibnamefont
  {Sch{\"u}ttler}},\ and\ \bibinfo {author} {\bibfnamefont {N.~E.}\
  \bibnamefont {Bickers}},\ }\bibfield  {title} {\bibinfo {title} {{$d$}-{{Wave
  Pairing}} in the {{Presence}} of {{Long-Range Coulomb Interactions}}},\
  }\href {https://doi.org/10.1103/PhysRevLett.82.1217} {\bibfield  {journal}
  {\bibinfo  {journal} {Phys. Rev. Lett.}\ }\textbf {\bibinfo {volume} {82}},\
  \bibinfo {pages} {1217} (\bibinfo {year} {1999})}\BibitemShut {NoStop}%
\bibitem [{\citenamefont {Braz}\ \emph
  {et~al.}(2024{\natexlab{a}})\citenamefont {Braz}, \citenamefont {Martins},\
  and\ \citenamefont {Dias~da Silva}}]{Braz:Phys.Rev.B:184502:2024}%
  \BibitemOpen
  \bibfield  {author} {\bibinfo {author} {\bibfnamefont {L.~B.}\ \bibnamefont
  {Braz}}, \bibinfo {author} {\bibfnamefont {G.~B.}\ \bibnamefont {Martins}},\
  and\ \bibinfo {author} {\bibfnamefont {L.~G. G.~V.}\ \bibnamefont {Dias~da
  Silva}},\ }\bibfield  {title} {\bibinfo {title} {Superconductivity from spin
  fluctuations and long-range interactions in magic-angle twisted bilayer
  graphene},\ }\href {https://doi.org/10.1103/PhysRevB.109.184502} {\bibfield
  {journal} {\bibinfo  {journal} {Phys. Rev. B}\ }\textbf {\bibinfo {volume}
  {109}},\ \bibinfo {pages} {184502} (\bibinfo {year}
  {2024}{\natexlab{a}})}\BibitemShut {NoStop}%
\bibitem [{\citenamefont {Xiao}\ \emph {et~al.}(2012)\citenamefont {Xiao},
  \citenamefont {Liu}, \citenamefont {Feng}, \citenamefont {Xu},\ and\
  \citenamefont {Yao}}]{Xiao_Phys.Rev.Lett._196802_2012}%
  \BibitemOpen
  \bibfield  {author} {\bibinfo {author} {\bibfnamefont {D.}~\bibnamefont
  {Xiao}}, \bibinfo {author} {\bibfnamefont {G.-B.}\ \bibnamefont {Liu}},
  \bibinfo {author} {\bibfnamefont {W.}~\bibnamefont {Feng}}, \bibinfo {author}
  {\bibfnamefont {X.}~\bibnamefont {Xu}},\ and\ \bibinfo {author}
  {\bibfnamefont {W.}~\bibnamefont {Yao}},\ }\bibfield  {title} {\bibinfo
  {title} {Coupled spin and valley physics in monolayers of
  ${\mathrm{mos}}_{2}$ and other group-vi dichalcogenides},\ }\href
  {https://doi.org/10.1103/PhysRevLett.108.196802} {\bibfield  {journal}
  {\bibinfo  {journal} {Phys. Rev. Lett.}\ }\textbf {\bibinfo {volume} {108}},\
  \bibinfo {pages} {196802} (\bibinfo {year} {2012})}\BibitemShut {NoStop}%
\bibitem [{\citenamefont {Parcollet}\ \emph {et~al.}(2015)\citenamefont
  {Parcollet}, \citenamefont {Ferrero}, \citenamefont {Ayral}, \citenamefont
  {Hafermann}, \citenamefont {Krivenko}, \citenamefont {Messio},\ and\
  \citenamefont {Seth}}]{ParcolletTRIQSToolboxForResearch2015}%
  \BibitemOpen
  \bibfield  {author} {\bibinfo {author} {\bibfnamefont {O.}~\bibnamefont
  {Parcollet}}, \bibinfo {author} {\bibfnamefont {M.}~\bibnamefont {Ferrero}},
  \bibinfo {author} {\bibfnamefont {T.}~\bibnamefont {Ayral}}, \bibinfo
  {author} {\bibfnamefont {H.}~\bibnamefont {Hafermann}}, \bibinfo {author}
  {\bibfnamefont {I.}~\bibnamefont {Krivenko}}, \bibinfo {author}
  {\bibfnamefont {L.}~\bibnamefont {Messio}},\ and\ \bibinfo {author}
  {\bibfnamefont {P.}~\bibnamefont {Seth}},\ }\bibfield  {title} {\bibinfo
  {title} {Triqs: A toolbox for research on interacting quantum systems},\
  }\href {https://doi.org/10.1016/j.cpc.2015.04.023} {\bibfield  {journal}
  {\bibinfo  {journal} {Comp. Phys. Comm.}\ }\textbf {\bibinfo {volume}
  {196}},\ \bibinfo {pages} {398–415} (\bibinfo {year} {2015})}\BibitemShut
  {NoStop}%
\bibitem [{\citenamefont {Strand}(2019)}]{StrandTPRFToolboxTRIQS2019}%
  \BibitemOpen
  \bibfield  {author} {\bibinfo {author} {\bibfnamefont {H.~U.~R.}\
  \bibnamefont {Strand}},\ }\href@noop {} {\bibinfo {title} {Two-particle
  response function toolbox (tprf) for triqs}},\ \bibinfo {howpublished}
  {\url{https://github.com/TRIQS/tprf}} (\bibinfo {year} {2019}),\ \bibinfo
  {note} {gitHub repository}\BibitemShut {NoStop}%
\bibitem [{\citenamefont {Kaye}\ \emph {et~al.}(2022)\citenamefont {Kaye},
  \citenamefont {Chen},\ and\ \citenamefont
  {Parcollet}}]{KayeDiscreteLehmannRepresentationImaginary2022}%
  \BibitemOpen
  \bibfield  {author} {\bibinfo {author} {\bibfnamefont {J.}~\bibnamefont
  {Kaye}}, \bibinfo {author} {\bibfnamefont {K.}~\bibnamefont {Chen}},\ and\
  \bibinfo {author} {\bibfnamefont {O.}~\bibnamefont {Parcollet}},\ }\bibfield
  {title} {\bibinfo {title} {Discrete lehmann representation of imaginary-time
  green’s functions},\ }\href {https://doi.org/10.1103/PhysRevB.105.235115}
  {\bibfield  {journal} {\bibinfo  {journal} {Phys. Rev. B}\ }\textbf {\bibinfo
  {volume} {105}},\ \bibinfo {pages} {235115} (\bibinfo {year}
  {2022})}\BibitemShut {NoStop}%
\bibitem [{\citenamefont {Rømer}\ \emph {et~al.}(2019)\citenamefont {Rømer},
  \citenamefont {Scherer}, \citenamefont {Eremin}, \citenamefont {Hirschfeld},\
  and\ \citenamefont
  {Andersen}}]{RoemerKnightShiftLeadingSuperconductingInstability2019}%
  \BibitemOpen
  \bibfield  {author} {\bibinfo {author} {\bibfnamefont {A.~T.}\ \bibnamefont
  {Rømer}}, \bibinfo {author} {\bibfnamefont {D.~D.}\ \bibnamefont {Scherer}},
  \bibinfo {author} {\bibfnamefont {I.~M.}\ \bibnamefont {Eremin}}, \bibinfo
  {author} {\bibfnamefont {P.~J.}\ \bibnamefont {Hirschfeld}},\ and\ \bibinfo
  {author} {\bibfnamefont {B.~M.}\ \bibnamefont {Andersen}},\ }\bibfield
  {title} {\bibinfo {title} {Knight shift and leading superconducting
  instability from spin fluctuations in {{Sr$_2$RuO$_4$}}},\ }\href
  {https://doi.org/10.1103/PhysRevLett.123.247001} {\bibfield  {journal}
  {\bibinfo  {journal} {Phys. Rev. Lett.}\ }\textbf {\bibinfo {volume} {123}},\
  \bibinfo {pages} {247001} (\bibinfo {year} {2019})}\BibitemShut {NoStop}%
\bibitem [{\citenamefont {Sakakibara}\ \emph {et~al.}(2012)\citenamefont
  {Sakakibara}, \citenamefont {Usui}, \citenamefont {Kuroki}, \citenamefont
  {Arita},\ and\ \citenamefont
  {Aoki}}]{sakakibaraOriginMaterialDependence2012}%
  \BibitemOpen
  \bibfield  {author} {\bibinfo {author} {\bibfnamefont {H.}~\bibnamefont
  {Sakakibara}}, \bibinfo {author} {\bibfnamefont {H.}~\bibnamefont {Usui}},
  \bibinfo {author} {\bibfnamefont {K.}~\bibnamefont {Kuroki}}, \bibinfo
  {author} {\bibfnamefont {R.}~\bibnamefont {Arita}},\ and\ \bibinfo {author}
  {\bibfnamefont {H.}~\bibnamefont {Aoki}},\ }\bibfield  {title} {\bibinfo
  {title} {Origin of the material dependence of $t_c$ in the single-layered
  cuprates},\ }\href {https://doi.org/10.1103/PhysRevB.85.064501} {\bibfield
  {journal} {\bibinfo  {journal} {Phys. Rev. B}\ }\textbf {\bibinfo {volume}
  {85}},\ \bibinfo {pages} {064501} (\bibinfo {year} {2012})}\BibitemShut
  {NoStop}%
\bibitem [{\citenamefont
  {Boehnke}(2015)}]{boehnkeSusceptibilitiesMaterialsMultiple2015}%
  \BibitemOpen
  \bibfield  {author} {\bibinfo {author} {\bibfnamefont {L.~V.}\ \bibnamefont
  {Boehnke}},\ }\emph {\bibinfo {title} {Susceptibilities in Materials with
  Multiple Strongly Correlated Orbitals}},\ \href@noop {} {Ph.D. thesis},\
  \bibinfo  {school} {Universit{\"a}t Hamburg}, \bibinfo {address} {Hamburg}
  (\bibinfo {year} {2015})\BibitemShut {NoStop}%
\bibitem [{\citenamefont {Braz}\ \emph
  {et~al.}(2024{\natexlab{b}})\citenamefont {Braz}, \citenamefont {Nag},\ and\
  \citenamefont
  {Black-Schaffer}}]{BrazCompetingMagneticStatesSurfaceMultilayer2024}%
  \BibitemOpen
  \bibfield  {author} {\bibinfo {author} {\bibfnamefont {L.~B.}\ \bibnamefont
  {Braz}}, \bibinfo {author} {\bibfnamefont {T.}~\bibnamefont {Nag}},\ and\
  \bibinfo {author} {\bibfnamefont {A.~M.}\ \bibnamefont {Black-Schaffer}},\
  }\bibfield  {title} {\bibinfo {title} {Competing magnetic states on the
  surface of multilayer abc-stacked graphene},\ }\href
  {https://doi.org/10.1103/PhysRevB.110.L241401} {\bibfield  {journal}
  {\bibinfo  {journal} {Phys. Rev. B}\ }\textbf {\bibinfo {volume} {110}},\
  \bibinfo {pages} {L241401} (\bibinfo {year}
  {2024}{\natexlab{b}})}\BibitemShut {NoStop}%
\bibitem [{\citenamefont {Yi}\ \emph {et~al.}(2019)\citenamefont {Yi},
  \citenamefont {Chen},\ and\ \citenamefont
  {Yu}}]{Yi_RecentAdvancesQuantumEffects2DMaterials_2019}%
  \BibitemOpen
  \bibfield  {author} {\bibinfo {author} {\bibfnamefont {Y.}~\bibnamefont
  {Yi}}, \bibinfo {author} {\bibfnamefont {Z.}~\bibnamefont {Chen}},\ and\
  \bibinfo {author} {\bibfnamefont {X.-F.}\ \bibnamefont {Yu}},\ }\bibfield
  {title} {\bibinfo {title} {Recent advances in quantum effects of 2d
  materials},\ }\href {https://doi.org/10.1002/qute.201800111} {\bibfield
  {journal} {\bibinfo  {journal} {Adv. Quantum Technol.}\ }\textbf {\bibinfo
  {volume} {2}},\ \bibinfo {pages} {1800111} (\bibinfo {year}
  {2019})}\BibitemShut {NoStop}%
\bibitem [{\citenamefont {Löthman}\ and\ \citenamefont
  {Black-Schaffer}(2017)}]{LöthmanUniversalPhaseDiagramsSuperconducting2017}%
  \BibitemOpen
  \bibfield  {author} {\bibinfo {author} {\bibfnamefont {T.}~\bibnamefont
  {Löthman}}\ and\ \bibinfo {author} {\bibfnamefont {A.~M.}\ \bibnamefont
  {Black-Schaffer}},\ }\bibfield  {title} {\bibinfo {title} {Universal phase
  diagrams with superconducting domes for electronic flat bands},\ }\href
  {https://doi.org/10.1103/PhysRevB.96.064505} {\bibfield  {journal} {\bibinfo
  {journal} {Phys. Rev. B}\ }\textbf {\bibinfo {volume} {96}},\ \bibinfo
  {pages} {064505} (\bibinfo {year} {2017})}\BibitemShut {NoStop}%
\bibitem [{\citenamefont {Barreto~Braz}\ and\ \citenamefont {Dias~da
  Silva}(2026)}]{ZenodoData_tWSe2}%
  \BibitemOpen
  \bibfield  {author} {\bibinfo {author} {\bibfnamefont {L.}~\bibnamefont
  {Barreto~Braz}}\ and\ \bibinfo {author} {\bibfnamefont {L.}~\bibnamefont
  {Dias~da Silva}},\ }\bibfield  {title} {\bibinfo {title} {Reentrant
  superconductivity and stoner boundaries in twisted wse$_2$},\ }\href
  {https://doi.org/https://doi.org/10.5281/zenodo.18461522}
  {https://doi.org/10.5281/zenodo.18461522} (\bibinfo {year}
  {2026})\BibitemShut {NoStop}%
\end{thebibliography}

%

\end{document}